\UseRawInputEncoding
\documentclass[aip,twocolumn,reprint,superscriptaddress,amsmath,amssymb,nofootinbib]{revtex4-1}
\usepackage{fullpage, amsmath, xparse, amssymb, mathtools, graphicx, braket,xcolor,verbatim}
\usepackage{natbib}
\usepackage{setspace}
\usepackage[ total={8in, 11in}, margin=0.9in]{geometry}
\usepackage[colorlinks=true,linkcolor=blue,citecolor=blue]{hyperref}
\makeatletter
\let\@currsize\normalsize
\makeatother

\def\changes{\textcolor{black}}
\def\changesNew{\textcolor{black}}

\usepackage{multirow}
\usepackage{cancel}
\usepackage[normalem]{ulem}
\usepackage{tikz} 
\usetikzlibrary{shapes,arrows,positioning,automata,backgrounds,calc,er,patterns,decorations.markings,snakes}

\tikzset{->-/.style={decoration={
  markings,
  mark=at position #1 with {\arrow[scale=2]{>}}},postaction={decorate}}}
 
\tikzset{-<-/.style={decoration={
  markings,
  mark=at position #1 with {\arrow[scale=2]{<}}},postaction={decorate}}}

\tikzset{
photon/.style={decorate, draw=black,
    decoration={coil,aspect=0}}
 }

\tikzset{
gluon/.style={decorate, draw=black,
    decoration={coil,amplitude=4pt, segment length=5pt}}
 }

\begin{document} 

\fontdimen2\font=0.5ex
\title{On inference of quantization from gravitationally induced entanglement}
\author{Vasileios Fragkos$^*$}
\email{vasileios.fragkos@fysik.su.se}

\affiliation{Department of Physics, Stockholm University, SE-106 91 Stockholm, Sweden}
\author{Michael Kopp$^*$}
\email{michael.kopp@su.se}
\affiliation{Department of Physics, Stockholm University, SE-106 91 Stockholm, Sweden}
\affiliation{Nordita,
KTH Royal Institute of Technology and Stockholm University,
Hannes Alfv\'ens v\"ag 12, SE-106 91 Stockholm, Sweden}
\author{Igor Pikovski} 
\email{igor.pikovski@fysik.su.se}
\affiliation{Department of Physics, Stockholm University, SE-106 91 Stockholm, Sweden}
\affiliation{Department of Physics, Stevens Institute of Technology, Hoboken, NJ 07030, USA }

\begin{abstract}
Observable signatures of the quantum nature of gravity at low energies have recently emerged as a promising new research field. 
One prominent avenue is to test for gravitationally induced entanglement between two mesoscopic masses prepared in spatial superposition.
Here we analyze such proposals and what one can infer from them about the quantum nature of gravity, as well as the electromagnetic analogues of such tests. We show that it is not possible to draw conclusions about mediators: even within relativistic physics, entanglement generation can equally be described in terms of mediators or in terms of non-local processes -- relativity does not dictate a local channel.
Such indirect tests therefore have limited ability to probe the nature of the process establishing the entanglement
as their interpretation is inherently ambiguous. 
We also show that cosmological observations already demonstrate some aspects of quantization that these proposals aim to test. Nevertheless, the proposed experiments would probe how gravity is sourced by spatial superpositions of matter, an untested new regime of quantum physics.

\end{abstract}

\maketitle

\section{Introduction} \label{sec:intro}

Quantum gravity remains one of the main challenges of modern physics. Many promising theoretical avenues are being pursued \cite{green_schwarz_witten_2012,rovelli_2004} but to date 
no complete theory of quantum gravity is known. An alternative route has been proposed by Roger Penrose \cite{Penrose1996GReGr..28..581P,PENROSE1998RSPTA.356.1927P} and others \cite{diosi1987PhLA..120..377D,Diosi,CollapseRevModPhys.85.471} in which quantum mechanical unitarity breaks down at sufficiently large mass scales. 
Such proposals have ushered in the exciting new field of experimental searches for quantum gravity at low energies, as they provide distinct signatures that can be observed if sufficiently large systems can be controlled and prepared in quantum superpositions.\cite{bose1999scheme, PhysRevLett.91.130401, kleckner2008creating,carney2019tabletop}  These proposals and the rapid progress in quantum control of larger and larger systems \cite{fein2019quantum, delic2020cooling} has sparked an interest in experimental signatures of the interface between quantum mechanics and gravity at low energies both in terms of testing speculative new models \cite{Penrose1996GReGr..28..581P,PENROSE1998RSPTA.356.1927P, diosi1987PhLA..120..377D,Diosi,CollapseRevModPhys.85.471, pikovski2012probing, bekenstein2012tabletop, kafri2014classical,belenchia2016testing} and expected new signatures from gravitational and quantum physics as we know them. \cite{zych2011quantum,Pikovskitimedilation2015,RohlichFolmanclocks,brodutch2015post}
To date, no experiment exists that either proves any of the new physics or the quantum nature of gravitational degrees of freedom, and it also seems that Gedankenexperiments in support of quantization are inconclusive.  \cite{Dyson:2013hbl,Baym_2009,belenchia2018quantum,rydving2021gedanken} Thus, any experimental signature that would suggest the quantum nature of gravity would be of extreme interest and shed light on uncharted territory. 

\begin{figure}[t]
    \centering
    \includegraphics[scale=0.5]{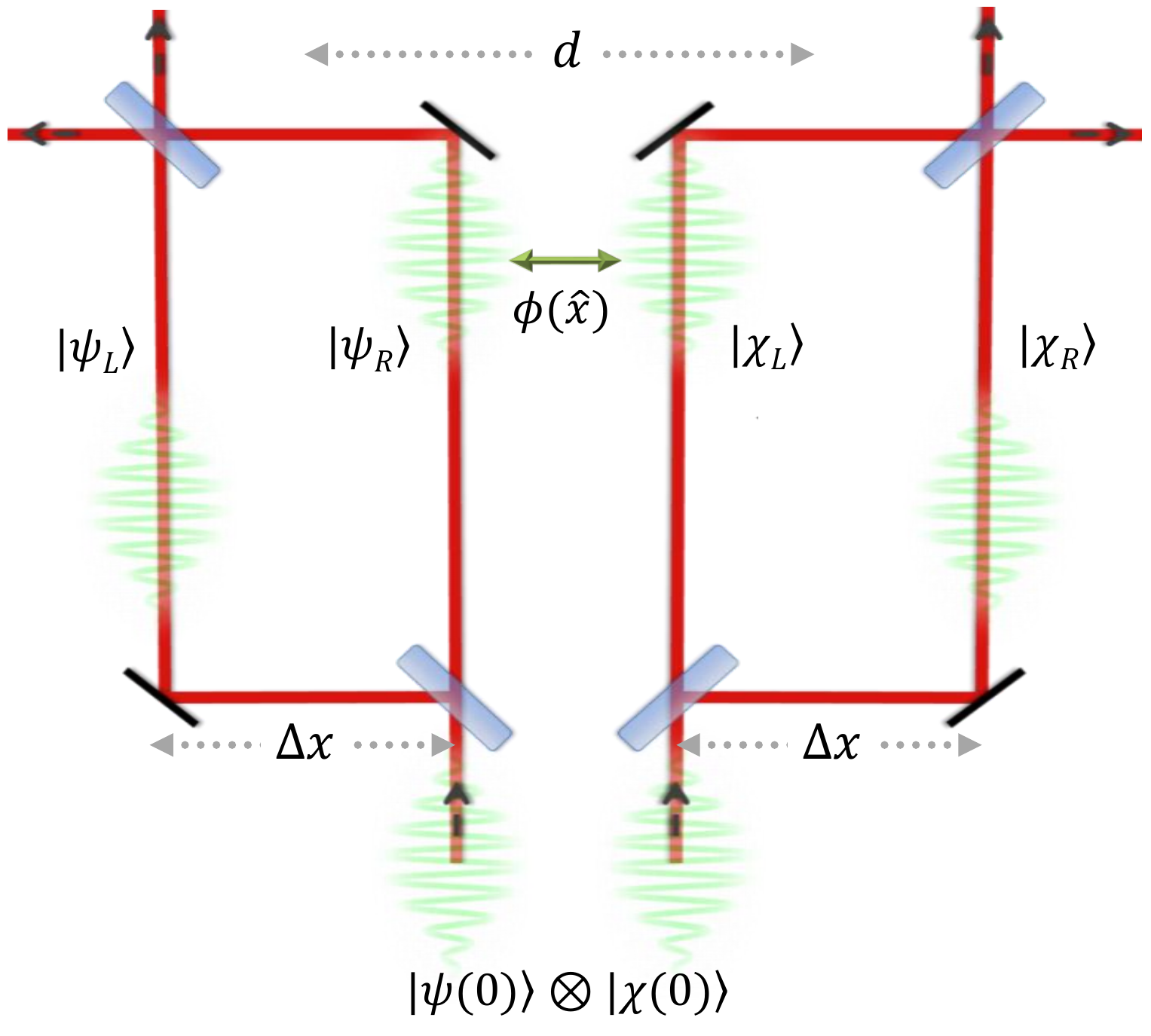}
    \caption{\small Schematic of an experimental setup to test for gravitationally induced entanglement. Two system are brought into a spatial superposition, denoted by $\ket{\Psi}$ and  $\ket{\chi}$, respectively. Due to their mutual, position-dependent interaction, here the Newtonian gravitational potential $\hat \Phi= -G m_1 m_2 /|\hat{x}_1 - \hat{x}_2|$, the two systems become entangled. The authors of Refs. \cite{BoseAnupam, MarlettoVedral2017} proposed such a setup to prove quantization of gravity, based on the argument that local, classical operations (LOCC) cannot generate entanglement. This conclusion, however, depends on what prior assumptions one is willing to take for granted. }
    \label{fig:setup}
\end{figure}

One promising experimental avenue that aims to infer the quantum nature of gravity is based on creating gravitationally induced entanglement between two source masses\cite{BoseAnupam,MarlettoVedral2017}  (GIE).\footnote{Some manuscripts refer to such setups as BMV, referring to some of the authors of the proposals.} 
In this variation of the well-known Colella-Overhauser-Werner setup,\cite{COWPhysRevLett.34.1472} two matter-wave interferometers are placed very close to each other (see Figure \ref{fig:setup}). The two interferometers create the superposition states $(\ket{\Psi_L} + \ket{\Psi_R})/\sqrt{2}$ and $(\ket{\chi_L} + \ket{\chi_R})/\sqrt{2}$, respectively, each of size $\Delta x$ and their centers a distance $d$ apart.
If the matter waves are sufficiently heavy they will source observable gravitational effects themselves, and the Newtonian interaction will generate entanglement between the two systems. After time $t$ the state becomes $(\ket{\Psi_L}\ket{\chi_L} + e^{i \varphi_+} \ket{\Psi_L}\ket{\chi_R} + e^{i \varphi_-} \ket{\Psi_R}\ket{\chi_L} + \ket{\Psi_R}\ket{\chi_R})/2  $ . The two relative phases are induced by the Newtonian interaction and are given by $\varphi_{\pm}=\frac{G m_1 m_2 t }{\hbar} \left( \frac{1}{d \pm \Delta x} - \frac{1}{d} \right)$. 
In this idealized case (neglecting noise) the state is entangled for any phases $\varphi_+ + \varphi_- \neq 2 \pi n$, $n \in \mathbb{Z}$, and maximally entangled for $\varphi_+ + \varphi_- = (2n +1) \pi$, which can be measured through an entanglement witness.\cite{BoseAnupam,chevalier2020witnessing, tilly2021qudits, guff2022optimal} 
The authors of Refs.\cite{BoseAnupam, MarlettoVedral2017}  argue that this entanglement generation is an indirect witness of the quantization of gravity. 
The essence of the argument is that the interaction is not instantaneous as suggested by the Newtonian description, but mediated by some local degrees of freedom, and since entanglement cannot be generated by local operations and classical communication (LOCC)\cite{PhysRevLett.66.1119,chitambar2014everything}, these degrees of freedom have to be quantized. 
Since the original proposals, several variations of this concept have been put forward. \cite{FOLMANmargalit2018realization,Bosefoll,followups,Pleniopedernales2021enhancing,marletto2020witnessing,kent2021testing,PRXQuantum.2.030330,ma2022limits,hosten2022constraints,streltsov2022significance}
However, the very question if such proposals can test the quantum nature of gravity has remained open for debate. Some have argued against it \cite{AnastopoulosHu2018,HallReginatto2018,hall2021comment} while others put arguments forward in favor. \cite{MarlettoVedral2018, Carney2021,belenchia2018quantum,followups, RovelliChristod2019,ChristodoulouEtal2022,danielson2022gravitationally} The main argument in favor of inferring the quantum nature of the mediating gravitational field stems from the quantum information perspective that limits entanglement generation from LOCC and the assumption that the interaction is induced by local mediators. In contrast, the major argument against the inference of quantum gravity is the use of the Newtonian limit in which the interaction is mediated non-locally. 

Here we clarify what conclusions one can draw from the experiment by critically exploring the assumptions that go into it. By definition, if entanglement is observed in the experiment, then this demonstrates a non-LOCC type channel. This alone is of course of great interest. Nevertheless, the open question around these proposals -- and what we address in this work -- is if one can learn more about the nature of the underlying system that mediates entanglement, namely of the gravitational field. As we will show below, the question remains ambiguous and depends on what prior assumptions one takes for granted -- in some cases even what metaphysical view one prefers. 
We show that while inferring the existence of gravitons is reasonable if one assumes the validity of relativistic field theory, this interpretation is not unique. It relies on the assumption of local mediators for entanglement, which in relativity are not the physical gravitons for this case, but particles that are otherwise unmeasurable. Moreover, models with inherent non-locality can reproduce the entanglement without the existence of any mediators. Importantly, while the term locality is often interchangeably used with the relativistic notion of Lorentz invariance, it needn't be: even within relativistic physics as we know it, entanglement generation can be described non-locally by eliminating redundant gauge degrees of freedom. And more generally, a fully non-local interpretation exists -- the Wheeler-Feynman absorber formulation.\cite{Tetrode1922,Lewis1926,Fokker1929,WheelerFeynman1945,WheelerFeynman1949} When applied to quantized charges \cite{HoyleNarlikar1969, HoyleNarlikar1971, Davies1971, Davies1972, Pegg1975, Pegg1979, Durrant1980, HoyleNarlikar1995} it reproduces all physical predictions of Quantum Electrodynamics (QED) including entanglement, but the theory does not contain quantized, local degrees of freedom that create the entanglement. If such an absorber theory can be formulated for gravity, it would explain gravitationally induced entanglement without introducing gravitons, while still being consistent with classical general relativity. With a viable non-local explanation for the generation of entanglement, its observation alone is therefore insufficient to infer the existence of quantized mediators. However, in contrast to the electromagnetic case, it remains an open question if an absorber-type formulation exists for full general relativity although it exists in the relativistic weak field  limit. Regardless of the question how entanglement is mediated, the experiment involves superpositions of source masses. As such, observation of entanglement thus supports the view that gravitational fields are sourced coherently by a superposition of sources, if quantum physics is not modified. This is independent of the question of what mediates entanglement and does not rely on the LOCC argument. The experiment, if successful, would prove this superposition of Newtonian fields, without the above mentioned ambiguities, as also highlighted in Refs.  \cite{RovelliChristod2019,ChristodoulouEtal2022}. Nonetheless, we argue that it can already be inferred from cosmological observations that quantum states coherently source Newtonian fields.

In this manuscript we explore systematically what conclusions one can draw from a GIE experiment, depending on what prior assumptions one takes for granted.  Our work is structured as follows: In Section \ref{sec-NR} we discuss the most conservative case, the Newtonian limit without any additional assumptions. In Section \ref{sec-EM} we focus on QED as an analogous case, and show that the conclusions are ambiguous and depend on (equivalent) formulations of the field theory. Section \ref{sec-GR} discusses the case of relativistic weak-field quantum gravity relevant to the GIE experiment.  Section \ref{sec-path} revisits all previously discussed formulations from the perspective of path integral quantisation, providing a unified picture. 
We focus in section \ref{sec:cosmo} on cosmology and show that current observations already give evidence of quantum weak-field gravity as in the GIE proposal. Our discussion in Section \ref{sec-discussion} gives a brief comparison to other works and provides an overview of the different interpretations and conclusions that one can draw from GIE.


\section{Non-relativistic case} \label{sec-NR}
We first focus on the physics in the non-relativistic limit, which is entirely sufficient to explain the setup. At this level, there is no assumption of any underlying field theory or additional ontology. It is the  minimal interpretation, as no additional assumptions about the underlying physics beyond what is experimentally measured have to be imposed.

In the non-relativistic limit there is a direct analogy between the Newtonian gravitational interaction and the electrostatic Coulomb law. To clarify the relevant physics, we briefly review how these enter the quantum theory of non-relativistic particles.
This will highlight the non-locality of the non-relativistic theory, and to what extent one can say that the Coulomb and Newtonian potentials ``are quantised''.

The non-relativistic quantum mechanics of $N=2$ interacting charged particles of charge $q$ and mass $m$ is given by the Hamiltonian
\begin{align}
\label{HamiltonianElectrostatic}
    \hat H  &=  \sum_{i=1}^N\left(\frac{(\hat{\vec p}_i)^2}{2m}+ \frac{q^2}{8\pi \epsilon_0 }\sum_{j=1, j\neq i}^N\frac{1}{|\hat{\vec x}_i - \hat{\vec x}_j |} \right) \,.
 \end{align}
For opposite charges this comprises the simplest (``textbook'') model of the hydrogen atom. The analogous Hamiltonian of gravitationally interacting particles is given by  
\begin{align} \label{HamiltonianNewton}
    \hat H =  \sum_{i=1}^N\left(\frac{(\hat{\vec p}_i)^2}{2m}- \frac{m^2 G}{2 }\sum_{j=1, j\neq i}^N\frac{1}{|\hat{\vec x}_i - \hat{\vec x}_j |} \right) \,.
\end{align}
The latter Hamiltonian is sufficient to fully account for the proposed GIE experiments and is used to derive all the results in the proposals.\cite{BoseAnupam, MarlettoVedral2017}
Since \eqref{HamiltonianNewton} does not contain any quantized (or unquantized for that matter) gravitational degrees of freedom, this theory of gravitionally interacting particles has no Hilbert space associated with gravity.

It is instructive to reformulate the quantum mechanical $N$-body problem as a quantum field theory. 
This is sometimes known as ``second quantised'' theory, but it is fully equivalent to the conventional formulation in terms of particle position operators. 
The quantum field (operator) in the Schr\"odinger picture $ \hat \psi(\vec x) $ has to property  $\hat \psi^\dagger(\vec x_1) |0 \rangle = | \vec x_1 \rangle$, that is, it creates a particle at the position $\vec x_1$. 
A general $N$-particle state is constructed by acting with $\int d^3 x_1 ... d^3 x_N \Psi(\vec x_1, ..., \vec x_N) \hat \psi^\dagger(\vec x_1)...\psi^\dagger(\vec x_N)$ on the vacuum state, where $\Psi$ is the $N$-particle wave function.
The Hamiltonian \eqref{HamiltonianNewton} then takes the equivalent form
\begin{multline}
   \hat H = \int d^3 x\frac{\hbar^2}{2 m} \nabla \hat \psi^\dagger(\vec x) \nabla \hat \psi(\vec x)   - \\
    -\frac{ m^2 G }{2 } \int d^3 x d^3 x' \frac{  \hat \psi^\dagger(\vec x)\hat \psi(\vec x) \hat \psi^\dagger(\vec x')\hat \psi(\vec x') }{|\vec x-\vec{x}'|}\,,
\end{multline}
and analogously for the Coulomb case.
We can rewrite this Hamiltonian in the following way
\begin{equation}
 \hat H = \int d^3 x\Big( \frac{\hbar^2}{2 m} \nabla \hat \psi^\dagger(\vec x) \nabla \hat \psi(\vec x)   + 
    m \hat \Phi(\vec x) \hat \psi^\dagger(\vec x) \hat \psi(\vec x) \Big) 
\end{equation}
where the Newtonian potential satisfies
\begin{equation}
      \nabla^2 \hat \Phi(\vec x) = 4 \pi G m \hat \psi^\dagger(\vec x) \hat \psi(\vec x) \label{NewtonQuantised}\,.
\end{equation}
Thus, non-relativistic quantum mechanics implies that the Newtonian gravitational field is quantized, in the sense that it is a field operator. There is no option to treat $\hat \Phi(\vec x)$ as a c-number field 
since the right hand side of the defining equation of $\hat \Phi(\vec x)$ is an operator. 
One way to insist on the classicality of the potential is to modify this equation to $\nabla^2  \Phi(\vec x) = 4 \pi G m \langle \Psi(t) | \hat \psi^\dagger(\vec x) \hat \psi(\vec x)| \Psi(t) \rangle,$ 
which is the non-relativistic limit of the so-called semi-classical gravity.\cite{singh1989notes} This theory is expected to be incorrect \cite{PagePhysRevLett.47.979} in situations where the quantum state of matter corresponds to density distribution in large spatial superpositions, where a mean field description in terms of $\langle \Psi(t) | \hat \psi^\dagger(\vec x) \hat \psi(\vec x)| \Psi(t) \rangle$ does not capture the quantum state $| \Psi(t) \rangle$. More importantly, though, this theory is not quantum mechanics since the Schr\"odinger equation is modified. 
Any proposed theory in which the Newtonian potential field is a c-number field (and thus classical in this sense), is not equivalent to ``textbook'' quantum mechanics, and thus would predict different outcomes for the GIE setup. Conversely, standard quantum mechanics implies that the Newtonian gravitational potential field is quantized in the sense that it is a field operator
\begin{align} \label{eq:NewtonField}
    \hat \Phi(\vec x, t) =  -   m G \int d^3 x' \frac{ \hat \psi^\dagger(\vec x')\hat \psi(\vec x') }{|\vec x-\vec{x}'|} \,.
\end{align}
The quantisation of the source masses implies the operator nature of the potential.
Only in situations where a mean field description for the sources is adequate can the operator nature of $\hat \Phi$  be neglected 
(such as for quantum test bodies falling evolving in the gravitational field of Earth \cite{COWPhysRevLett.34.1472,zych2011quantum,Pikovskitimedilation2015}). 

The novelty of the GIE experiment is thus that the Newtonian field as in eq. \eqref{eq:NewtonField} has to be treated quantum mechanically, and cannot be approximated by a mean field description. The sources for gravity are themselves in superposition. While this regime of physics has never been directly tested, the predictions assume regular quantum mechanics \eqref{HamiltonianNewton}. 
The gravitationally active density is in a spatial superposition and thus the operator nature of the potential cannot be neglected.
However, nothing in this description involves mediators and thus nothing can be said whether gravitons are quantized, at this level of the description. 
Such claims rely on extra assumptions, such as the assumption that $\hat \Phi(\vec x)$ is the component of a more fundamental relativistic quantum field that has local degrees of freedom.

The following sections will answer under which conditions/assumptions our quantized Newtonian potential also implies the existence of quantized gravitons, and deals with questions of non-locality using the simpler but very analogous example of electromagnetism \changes{ (summarized in Table \ref{LOCCtable})}.

\section{Electromagnetic Field Theory} \label{sec-EM}
We now proceed to analyze possible conclusions beyond the Newtonian limit, under the assumption that one takes relativistic field theory for granted. Again, we note that the non-relativistic limit is sufficient to describe the experiment as in the section above. Now, however, we take the view that from other experiments and established physics we can additionally assume that there is an underlying field theory that mediates the entanglement, even if it is not directly probed in the experiment. The question is, can one conclude something more than what has been established in the previous section?

To simplify the discussion and gain intuition we will assume that the entanglement production in a GIE-like experiment is solely due to electric charge (thus technically it is electromagnetically induced entanglement), before discussing the case of gravity in the next section.
Our discussion within this section is then guaranteed to be based on established physics. 
Our task is thus to find out what we would learn about the quantum nature of electromagnetism in the hypothetical scenario where we only had the electromagnetic analogue of the GIE experiment at our disposal. In particular, we investigate which assumptions about electromagnetism are needed to infer the existence of (quantized) photons from witnessing entanglement from this experiment alone.

In this section we therefore slightly modify the experiment shown Fig.\,\ref{fig:setup} and assume that two objects that are put into superposition have electric charge $q$ with $q^2/\epsilon_0 \gg G m^2$, and interact via the Coulomb potential $\Phi_{em}$.


\begin{table}[h]
    \centering
    {\begin{tabular}{l |c | c}
        Formulation of Interaction & Equation & LOCC?\\ \hline \hline
        non-relativistic Coulomb &  \eqref{HamiltonianElectrostatic} & \cancel{L}OCC \\ \hline
         QED (Lorentz gauge)  & \eqref{Fullqedhamlorentz} &  LO\cancel{C}C\\ \hline
         QED (Coulomb gauge)  & \eqref{qedcoulombham} &  \cancel{L}OCC\\ \hline
        QED (Absorber, no EM field)  & \eqref{FullEM-WF} & \cancel{L}OCC \\ \hline \hline
      
    \end{tabular}
    }
    \caption{\small  An LOCC channel cannot generate entanglement. But if entanglement is observed, can one deduce more about the physics of the underlying process, such as quantized mediators? Comparison of quantum theories (first column)  of electrically charged particles and their defining equations (second column), that are expected to produce identical entanglement, and what aspect of LOCC they violate (third column) such that they can generate entanglement. Only the formulation of QED in the (local) Lorentz gauge includes quantized mediators relevant for the electromagnetic analog of the GIE experiment. The other formulations of QED remain equally viable, from which no quantization of the mediators can be inferred  from the LOCC argument, and the experiment cannot discern them. Even beyond this ambiguity related to the gauge freedom,
    the absorber formulation of QED by Hoyle, Narlikar\cite{HoyleNarlikar1995} and Davies\cite{Davies1971, Davies1972} (based on quantizing the classical Wheeler-Feynman  formulation of electrodynamics)  is experimentally indistinguishable from standard QED even in the fully relativistic regime. Thus the GIE experiment if applied to the electromagnetic interaction would not be able to tell us whether there is a Hilbert space associated with the electromagnetic field or not.
    As non-local entanglement generation cannot be ruled out, we therefore cannot conclude \changes{based on the LOCC argument} whether underlying mediators are quantized.}
    \label{LOCCtable}
\end{table}

\subsection{Local mediators: Lorentz invariant formulation of QED} \label{sec:EM-Lorentz}
The dynamics of the photon field in any Lorentz invariant gauge is local, as captured by the local interaction Lagrangian
\begin{equation} \label{EMint}
    \mathcal{L}_{int}= j^{\mu}(x) A_{\mu}(x) \,,
\end{equation}
where $A_{\mu}(x)=(\Phi_{em},\vec{A})$ is the electromagnetic four-potential,  $j_{\mu}(x)=(c\rho,\vec{j})$ the four-current and $x=(t,\vec x)$. In the following we set $c=1$. By choosing a particular gauge, namely the Lorentz gauge, locality becomes manifest even on the level of the mediators themselves, giving a very intuitive physical picture.

In this formulation of QED,\cite{Gupta:1949rh,Bleuler:1950cy} all four components of the gauge field $\hat{A}_{\mu}$ 
are quantized. Therefore, the theory contains 4 types of photons, each for every component of the four-potential. We expand the gauge field $\hat{A}_{\mu}$ in terms of creation and annihilation operators
\begin{multline} 
    \hat{A}_{\mu}(\vec{r})=\int d^3k\sqrt{\frac{ \hbar }{2\epsilon_0(2\pi)^3\omega_k}} \biggl[\hat{a}_{\mu}(\vec{k})e^{i\vec{k}\cdot\vec{r}}+\\+\hat{a}_{\mu}^{\dagger}(\vec{k})e^{-i\vec{k}\cdot\vec{r}}\biggr]
\end{multline}
where $\mu=(0,1,2,3)$.
The expression for the Hamiltonian is\cite{Cohen-TannoudjiEtal1992}
\begin{subequations}\label{Fullqedhamlorentz}
\begin{equation}
    \hat{H}=\hat{H}_P +\hat{H}_R+\hat{H}_I
\end{equation}
where $\hat{H}_P$ is the free particle Hamiltonian,  
\begin{multline}\label{qedhamlorentz}
 \hat{H}_R=\int d^3k \hbar\omega\biggl[\sum_{\lambda=\pm,l}\biggl(\hat{a}^{\dagger}_{\lambda}(\vec{k})\hat{a}_{\lambda}(\vec{k})\biggr)-\\-\hat{a}^{\dagger}_{s}(\vec{k})\hat{a}_{s}(\vec{k})+1\biggr]
\end{multline}
\end{subequations}
is the field Hamiltonian, which contains creation and annihilation operators for scalar (s), longitudinal (l) and transverse ($\pm$) photons.
 The interaction Hamiltonian between the particles and the field is
\begin{equation}\label{LorentzHamiltonian}
    \hat{H}_I=\int d^3r\bigl[-\hat{\vec{j}}(\vec{r})\cdot\hat{\vec{A}}(\vec{r})+\hat{\rho}(\vec{r})\hat{\Phi}_{em}(\vec{r})\bigr]\,.
\end{equation}

Note that in \eqref{qedhamlorentz}, the term which corresponds to scalar photons comes with a minus sign. Quantization of QED in Lorentz gauge, requires the use of the so-called indefinite Hilbert space metric. \cite{Gupta:1949rh} A pedagogical presentation of the method can be found in Ref. \cite{Cohen-TannoudjiEtal1992}.

As seen in eq.\,\eqref{qedhamlorentz}, this method of quantization introduces fictitious longitudinal and scalar photons that are not directly observable in any experiment (in the sense that expectation values of any observable, as well as in- and out-states in scattering theory can only involve quantum states where the scalar and longitudinal photons are in the vacuum state).
In presence of matter sources, 
scalar and longitudinal photons can appear in virtual states \cite{Franson2011} but have no observable consequences. Only the transverse photons are physical. Thus a subsidiary condition on the Hilbert space of physical states is imposed \cite{Gupta:1949rh}
\begin{equation}\label{subsidiary}
    \bigl(\hat{a}_l(\vec{k})-\hat{a}_s(\vec{k})+\hat{\lambda}(\vec{k})
\bigr)|\chi\rangle=0 
\end{equation}
$\forall \vec{k}$ and where $\hat{\lambda}(\vec{k})=\frac{1}{\omega\sqrt{2\epsilon_0\hbar\omega}}\hat{\rho}(\vec{k})$.\\

In the free theory we have $\hat{\lambda}(\vec{k})=0$ so the condition \eqref{subsidiary} takes the form
\begin{equation} \label{eq:LSphotons}
    \bigl(\hat{a}_l(\vec{k})-\hat{a}_s(\vec{k})\bigr)|\chi\rangle=0 
\end{equation} 
$\forall \vec{k}$.
This relation guarantees that, for every physical state $|\chi\rangle$, the effect of longitudinal and scalar photons cancel.\footnote{One can formally show, by making use of the so-called Ward identity, that the effect of scalar and longitudinal photons can be neglected in any QED scattering process. For a standard treatment, the reader is referred to Ref.\cite{Peskin:1995ev}.  } 

What is crucial is that entanglement is generated in a local and causal way, but it is mediated not by the physical photons but by the scalar and longitudinal ones.\cite{Franson2011}
This can be easily seen in a toy example (analogous to the GIE setup) of a system of two harmonic oscillators $(A,B)$ which are initially in a product state 
\begin{equation}\label{groundstate}
 |\Psi_0\rangle \equiv|0\rangle_A\otimes|0\rangle_B\otimes|\vec{0}\rangle_{\gamma}   
\end{equation}
 where $|0\rangle_i$ for $i=(A,B)$, corresponds to the ground state of the oscillators and $|\vec{0}\rangle_{\gamma}$ is a state which contains no photons. Coupling the system of the two oscillators to the electromagnetic field via the interaction Hamiltonian \eqref{LorentzHamiltonian}, the perturbed state can be calculated using time independent second order perturbation theory and it gets entangled between $A$ and $B$ \cite{Franson2011}
\begin{equation}\label{entangledstate}
    |\Psi\rangle \simeq |\Psi_0\rangle+\varepsilon_L|\Psi_2\rangle
\end{equation}
where
\begin{equation}\label{excitedstate}
    |\Psi_2\rangle =  |1\rangle_A\otimes|1\rangle_B\otimes|\vec{0}\rangle_{\gamma}
\end{equation}
corresponds to the second order $(\mathcal{O}(q^2))$ change in $|\Psi\rangle_0$ and $\varepsilon_L$ is given by 
\begin{equation}\label{perturbedstate}
    \varepsilon_L=\sum_{l}\frac{\langle \Psi_2|\hat{H}_I|l\rangle \langle l|\hat{H}_I|\Psi_0\rangle}{(E_0-E_1)(E_0-E_l)}\,.
\end{equation}
We have used a compact notation for the intermediate states $|l\rangle\equiv|l\rangle_A\otimes|l\rangle_B\otimes|\vec{k}\rangle_{\gamma}$\footnote{In principle, the intermediate photon states $|\vec{r}\rangle_{\gamma}$ contain scalar, longitudinal and transverse photons. For the electromagnetic analogue of the GIE proposal, we omit the contribution of transverse photons since the latter play no role in establishing the entanglement.}. $E_0$ corresponds to energy of the state \eqref{groundstate}, where the oscillators $(A,B)$ are in the ground state whereas $E_1$ is the energy of the perturbed state \eqref{excitedstate} where the oscillators occupy the first excited state $|1\rangle_{(A,B)}$. Both states contain no photons. On the other hand, $E_l$ corresponds to the energy of the intermediate states $|l\rangle$, where virtual photons do exist. It is important to notice that since $E_l\equiv E_{({l_A,l_B,l_{\gamma}})}=E_{l_A}+E_{l_B}+E_{l_{\gamma}}$ appears in the denominator of \eqref{perturbedstate}, one cannot trivially perform the summation over the oscillator states. 

The state \eqref{entangledstate} corresponds to an entangled state of the two oscillators $(A,B)$ for $\varepsilon_L \neq 0$. It's crucial that in \eqref{perturbedstate}, apart from summing over intermediate oscillator states, there is a summation over intermediate photon states $|\vec{k}\rangle_{\gamma}$. In a recent work, a similar computation was performed for the gravitational case but the summation over intermediate oscillator states has been omitted.\cite{bose2022mechanism} Thus we conclude that in this picture, the intermediate photons establish the entanglement between the two oscillators. In the near field regime in which the transverse photons do not contribute, the oscillators get entangled due to the exchange of scalar and longitudinal photons -- mediators that cannot be directly observed.

When using a local gauge such as the Lorentz gauge as the underlying ontology, the LOCC argument therefore implies that entanglement is created by quantized mediators, since locality is pre-assumed and thus classicality of the channel cannot hold if entanglement is generated. The  process is a particular combination of the scalar and longitudinal photons and does not directly reveal the quantum nature of transverse (physical) photons. If  QED was not pre-assumed, but instead, we attempted to infer the quantum nature of the electromagnetic field from the GIE experiment and the LOCC argument, we would still not be able to directly infer quantization of physically measurable mediators. However, consistency with Lorentz invariance and locality of the quantum fields will then also require these transverse degrees of freedom to exist and be quantized. This is a matter of theoretical consistency, but not something that can be directly inferred from the entanglement experiment, even in this picture. The experiment is not sensitive to actual physical mediators -- photons (or gravitons in the GIE case) -- and entanglement is created by other local mediators that can never be directly observed.  Nevertheless LOCC then implies that these must be quantized. 

\subsection{No mediators: QED in Coulomb gauge} \label{sec:CoulombQEDFranson}
We now turn to a different, but physically equivalent description of the process. The fundamental locality of eq. \eqref{EMint} does not necessarily imply locality of mediators of the interaction. Contrary to the covariant quantization of QED,\cite{Gupta:1949rh,Bleuler:1950cy} in the Coulomb gauge no scalar and longitudinal photons appear at any stage. The expression for the QED Hamiltonian in the Coulomb gauge  (assuming spinless particles for simplicity) is \cite{Cohen-TannoudjiEtal1992}
\begin{equation}\label{qedcoulombham}
\begin{split}
 \hat{H}  =\sum_{i}\frac{\bigl(\hat{\vec{p}}_i\bigr)^2}{2m_i}&+\int d^3r\hat{\vec{j}}(\vec{r})\cdot\hat{\vec{A}}_{\perp}(\vec{r})+\\&+\frac{1}{8\pi\epsilon_0}\int \int d^3 rd^3r'\frac{\hat{\rho}(\vec{r})\hat{\rho}(\vec{r}')}{|\vec{r}-\vec{r}'|}+\\ &+ \int d^3k\hbar\omega\sum_{\lambda=\pm}\bigl(\hat{a}^{\dagger}_{\lambda}(\vec{k})\hat{a}_{\lambda}(\vec{k})+\frac{1}{2}\bigr)
\end{split}
\end{equation}
where the first line describes the kinetic energy of particles as well as their interaction with the field, the second term is the Coulomb interaction energy and the third term corresponds to the the Hamiltonian of the radiation field. The operators $\hat{a}^{\dagger}_{\lambda}(\vec{k})$ and $\hat{a}_{\lambda}(\vec{k})$ create and annihilate only transverse (i.e. physical) photons with momentum $\vec{k}$ whereas $\hat{\vec{r}}_i$ and $\hat{\vec{p}}_i$ correspond to the $i$'s particle position and momentum operators.

The Hilbert space of the whole system (particles+EM field) is $\mathcal{H}=\mathcal{H}_P \otimes\mathcal{H}_R$, where $\mathcal{H}_P$ is the state space of the particles in which $\hat{\vec{r}}_i$ and $\hat{\vec{p}}_i$ act and $\mathcal{H}_R$ corresponds to the state space of the photon-radiation field in which $\hat{a}^{\dagger}_{\pm}(\vec{k})$ and $\hat{a}_{\pm}(\vec{k})$ act. Note that predominantly the third term in \eqref{qedcoulombham} generates the direct interaction between the two non-relativistic and quasi-static systems, and it is fundamentally non-local. Since radiation (retardation effects) are not important in the GIE experiment, one has thus to conclude that, in the formulation of QED in the Coulomb gauge, the entanglement between the charges is established non-locally, via the Coulomb term (as also pointed out in Ref. \cite{AnastopoulosHu2018}). In this case, the LOCC argument alone to prove that the underlying process that generates entanglement is itself a quantum system does not apply
and nothing about the mediators can be inferred. For further details regarding QED in Coulomb gauge, the reader is referred to Appendix \ref{sec:AppendixQEDCoulomb}.

In the toy example of the two oscillators $(A,B)$, the entangled state is computed using first order time independent perturbation theory \cite{Franson2011}
\begin{equation}\label{entangledstatecoulomb}
    |\Psi\rangle=|\Psi_0\rangle+\varepsilon_C|\Psi_1\rangle
\end{equation}
where initially the oscillators are in the ground state
\begin{equation}
 |\Psi_0\rangle \equiv|0\rangle_A\otimes|0\rangle_B  
\end{equation}
and the perturbed state is
\begin{equation}
|\Psi_1\rangle \equiv|1\rangle_A\otimes|1\rangle_B  
\end{equation}
where $\varepsilon_C$ is given by
\begin{equation}\label{perturbedstatecoulomb}
   \varepsilon_C=\frac{\langle \Psi_1|\hat{H}_C|\Psi_0\rangle}{(E_0-E_1)}
\end{equation}
and 
\begin{equation}
    \hat{H}_C=\frac{1}{8\pi\epsilon_0}\int \int d^3 rd^3r'\frac{\hat{\rho}(\vec{r})\hat{\rho}(\vec{r}')}{|\vec{r}-\vec{r}'|}
\end{equation}
is the Coulomb interaction Hamiltonian. The state \eqref{entangledstatecoulomb} corresponds to an entangled state of the two oscillators. 
However, in  \eqref{perturbedstatecoulomb}, there are no intermediate or final photon states. Therefore, in that formulation one can explicitly see that nothing can be inferred about photons, physical or unphysical, mediating the entanglement. We thus conclude that depending on the preferred picture within the full scope of QED, the LOCC argument can equivalently imply either quantized mediators as in the previous section, or a non-local interaction. Locality is therefore not a necessary principle, but a choice. While the Coulomb formulation of QED is fundamentally non-local, causality is preserved and is the fundamental principle that holds in all formulations, as discussed in more detail in section \ref{sec:causal}. Note that manifest Lorentz invariance is not necessarily lost in a non-local formulation of QED, see for example Ref.\cite{PhysRev.125.2189} in which the electromagnetic field is quantised using the field strength tensor instead of the vector potential. It is thus clear that non-locality is not a gauge artefact but a property of any formulation of QED that only quantizes the physical degrees of freedom of the electromagnetic field.

In fact all interactions of the standard model and gravity are gauge theories and thus if the formulation of these theories is restricted to involve only the dynamics of the physical degrees of freedom all these theories are non-local (in that the Lagrangian density is not a sum of local products of operators). 
Only with the help of auxiliary gauge degrees of freedom are these theories local. 
It thus is a matter of taste if locality is pre-assumed or not, and thus whether the LOCC argument implies the existence of quantum mediators or not.

\subsection{No mediators: QED as absorber theory}
\label{sec:nomedQEDabsorber}
Above we discussed different interpretations of the experiment based on choosing different gauges. This ambiguity was also pointed out in Ref.\cite{AnastopoulosHu2018}. It shows that the notion of local mediators is not necessary to explain the experiment, and thus no conclusion about its quantization can be drawn (see also Table \ref{LOCCtable}).

However, there is a deeper issue in the context of electromagnetism. One general loophole to conclude that a relativistic theory of interacting particles has to involve exchange bosons is that one can remove the mediators from the theory. 
An example of such a drastic approach is provided by the Wheeler-Feynman (WF) absorber theory of electromagnetism.\cite{Schwarzschild1903,Tetrode1922,Fokker1929,WheelerFeynman1945,WheelerFeynman1949}
In the absorber theory, the interaction between charges is non-local, captured by the Lagrangian 
\begin{subequations}\label{FullEM-WF}
\begin{align}\label{EM-WF}
\mathcal{L}_{int} = j^{\mu}(x) \int d^4x'\, j_\mu(x') \delta_{\rm D}\big[ (x- x')^2\big]\\
\delta_{\rm D}\big[ (x- x')^2\big] = \frac{1}{2} \big(G_-(x-x')+G_+(x-x')\big)
\end{align}\\
where $(x- x')^2= c^2 (t-t')^2 - |\vec x - \vec x'|^2$ is the Minkowski distance between $x=(t, \vec x)$ and $x'=(t', \vec x')$,  $\delta_{\rm D}$ the Dirac delta function, $G_-$ is the retarded and $G_+$ the advanced Green's function of the d'Alembert operator $\Box$:
\begin{equation}
    G_{\pm} = |\vec{x} - \vec x'|^{-1} \delta_{\rm D}\big[ c(t-t') \pm |\vec{x} - \vec x '| \big] \, . 
\end{equation}
\end{subequations}
The Lagrangian density \eqref{EM-WF} is manifestly causal due to $\delta_{\rm D}\big[ (x- x')^2\big]$ restricting the spatial non-locality to the future and past lightcone. \changes{It is manifestly non-local because the delta function has in general support arbitrarily far away from the position $\vec x$.}
Despite the intrinsic non-locality and seemingly drastic departure from our usual intuition for physics, it was shown that all results of classical electromagnetism, in particular Dirac's half-advanced-half-retarded solution to the radiation reaction problem, can be derived from such a theory if suitable absorbing boundary conditions are chosen. Within standard electromagnetism this corresponds to the assumption that all emitted radiation is eventually completely absorbed by some charge in the future. When and how this absorption happens is irrelevant. These boundary conditions are thus cosmological in nature and might appear contrived or unintuitive. But this is just a matter of taste and cannot be refuted logically or by any  observation -- it is just a reformulation of regular electromagnetism but with a different ontology.\footnote{In Sec.\,\ref{sec:grav-absorber} we discuss that in the case of non-perturbative gravity, an absorber theory is likely different from the conventional field theory formulation. However, in the weak-field regime relevant to GIE, gravitational absorber theory and standard weak-field GR are reformulations of each other. }
Furthermore, all results of QED can be recovered in such a theory as well by simply quantizing the charges, thereby obtaining a viable reformulation of QED without any electromagnetic degrees of freedom, as first shown by Hoyle and Narlikar in 1969 and further worked out by Davies, Pegg and others. \cite{HoyleNarlikar1969, HoyleNarlikar1971, Davies1971, Davies1972, Pegg1975, Pegg1979, Durrant1980, HoyleNarlikar1995}$^{,}$\footnote{
While all predictions for experimental outcomes seem to be the same for the Hoyle-Narlikar-Davies quantized version of the absorber theory, it should be pointed out that only path integral quantisation has been possible so far. Canonical quantisation is complicated by the fact that the Lagrangian contains a double integral over time. See however Ref.\cite{Lienert2018} for a canonical approach.}

In particular, Feynman suggested in 1948 \cite{RevModPhys.20.367} that one can replace \eqref{EMint} with \eqref{EM-WF} for any QED process which contains no external, real photons, without any change in the experimental results.  According to this theorem (see Ref.\cite{Davies1971} for a proof), one can eliminate all virtual photons from the mathematical description of the QED processes. This theorem directly applies to the EM analogue of the GIE proposal in which on-shell photons are not involved.\footnote{Note the there is technically a difference between an ``absorber theory of radiation'', \cite{WheelerFeynman1945} in which fields are an element of the theory but absorbing boundary conditions eliminate free fields, and an action-at-a-distance theory in which absorbing boundary conditions need to be assumed for the theory to be phenomenologically viable.\cite{WheelerFeynman1949} In this paper we lump together these two two types of absorber theories since they make the same predictions for the electromagnetic case, and more importantly for our discussion, upon quantisation photons do not exist in physical states in both these theories. }

While the physical predictions are the same, the interpretation of such a theory is different. In this action-at-a-distance formulation the electromagnetic field does not exist and particles interact directly, without mediators, on the lightcone, see \eqref{EM-WF}.
Such an ontology in the context of the GIE experiment has profound consequences: Since the direct interaction on the lightcone is non-local in space, the LOCC argument does not imply a quantised mediating system for entanglement.
This means that witnessing entanglement in the electromagnetic version of the GIE experiment could not  unambiguously tell us if the electromagnetic field needs to be quantised.
Thus entanglement alone only proves source mass quantisation, since there is an experimentally equivalent alternative formulation, where no mediators are involved in producing the entanglement.

In conclusion, the electromagnetic analogue of the GIE experiment would tell us nothing about mediators beyond our prior believes about which formulation of electromagnetism is more likely to be true. One either believes that physics is local which means firstly that longitudinal and scalar photons establish the entanglement -- and not the physical transverse photons --, and secondly that transverse photons must exist and be quantized as well, since only the combination of these three types of photons can build a local field theory. Or one doesn't manually exclude
non-local establishment of entanglement, which is the case in the  Coulomb formulation of QED.
Alternatively, in an absorber framework only the sources are real and the entire electromagnetic field is a mathematical fiction and should not be quantized.
The entanglement in this picture does not stem from mediators but from the charges interacting non-locally. All formulations, however, yield the exact same experimental outcomes.
\changes{It is important to note that the case of \cancel{L}OCC (where locality does not hold, see Table \ref{LOCCtable}) does not necessarily imply a classical channel. It simply means that a \changesNew{local} quantum channel is not necessary to produce the entanglement.
In the case of non-local formulations of QED, entanglement is produced through a quantum channel which can be directly inferred from the Hamiltonian \eqref{qedcoulombham} or Lagrangian \eqref{FullEM-WF}. But in the case of Coulomb gauge or absorber theory that quantum channel is not a mediating degree of freedom but a non-local interaction between the charges.}

\subsection{Locality vs. Causality} \label{sec:causal}

As we have discussed above, locality is not a necessary condition even in a relativistic theory. One is free to choose a local and manifestly covariant formulation of QED, but a non-local description is also always possible. Nevertheless causality is preserved in all of the formulations. There is thus a fundamental difference between non-locality and causality, and only the latter is fundamentally required by Lorentz invariance.

In the Coulomb gauge, despite the various non-local (instantaneous) pieces that appear in \eqref{qedcoulombham}, only causal physical processes are described.\cite{Rohrlich2002,PhysRevA.38.4897} In particular, it was shown in Refs. \cite{Jackson2002,PhysRevA.38.4897} that, in Coulomb gauge, the instantaneous Coulomb potential cancels exactly the instantaneous piece that appears in the vector field $\hat{\vec{A}}_{\perp}$ to give rise to causal (retarded) EM fields, the so-called Jefimenko fields.\cite{Jefimenko1989}

From a Quantum field theory point of view, the photon propagator contains a piece which violates Lorentz invariance. This term in the propagator exactly cancels the acausal, Lorentz violating Coulomb potential that appears in the Hamiltonian \eqref{qedcoulombham}, to give a perfectly Lorentz invariant, causal QED. A detailed discussion can also be found in Ref.\cite{0521670535}.

The absorber theory is manifestly causal due to the lightcone integral in eq \eqref{EM-WF}. The absorbing boundary conditions are not required for causality, but they ensure approximate forward causation, see Ref.\cite{WheelerFeynman1945}.

Overall, since all these formulations are just different ways of interpreting Maxwell's equations, it is clear that relativistic causality is preserved in any physical process described in these frameworks. Nevertheless, even while being causal, a non-local ontology is admitted by relativity. 
This peculiarity is at the core of other seeming puzzles of classical electromagnetism, such as the question 
how the Jefimenko equations \cite{Jefimenko1989} for $\vec{E}$ and $\vec{B}$  can be causal despite the Coulomb equation, $\nabla \cdot \vec{E} = \rho/\epsilon_0$, suggesting the opposite.\cite{Wong2010, Jackson2002}
The Coulomb equation seems acausal because $\nabla \cdot \vec{E}$, and thus the longitudinal part of $\vec E$, depend instantaneously on the charge density $\rho$. 
To see explicitly that this instantaneous dependence is consistent with causality one can use the dynamical Maxwell equations to write $\Box (\nabla \cdot\vec{E}) = \frac{1}{\epsilon_0} \nabla^2 \rho + \mu_0 \nabla \partial_t \vec{J}$. All relativistic field theories with first class constraints have the property that instantaneous constraint equations are compatible with the causal evolution.

As already discussed, the GIE proposal as well as its EM analogue, heavily rely on the notion of locality, which lies at the core of the LOCC argument. In the preceding subsection, we have concluded that locality is not a prerequisite, instead it's a matter of choice to formulate physics in a local or non-local way. This discussion addresses  similar questions and shares some similarities with the correspondence between DeWitt and Aharonov and Bohm  \cite{PhysRev.125.2189,PhysRev.125.2192} (see also Ref.\cite{PhysRev.128.2832}) shortly after their seminal work on the ``Significance of the EM potentials in quantum theory''.\cite{PhysRev.115.485} In particular, in Ref.\cite{PhysRev.125.2189}, DeWitt asks:
\emph{Which is more significant, the fact that nonlocal formulations of causal theories exist which deal only with observables, or the fact that in all known cases local formulations in terms of potentials also exist?}
And continues with the following statement:
\emph{In a similar vein the author disagrees with the assertion of Aharonov and Bohm that quantum electrodynamics is ultimately determined by the requirement that it be expressible in a local form. QED is really determined by experiment.} 

This assessment aligns with our interpretation of the electromagnetic analogue of the GIE proposal. QED is determined by experiment and it cannot discern a local from a non-local formulation. The underlying ontology corresponds to a \emph{metaphysical statement} which by no means is tested. Thus, any formulation of QED, local or non-local are equally viable, as long as the experimental outcomes are indistinguishable. For the EM analogue of GIE,  this implies that one cannot \emph{unambiguously} infer the existence of local mediators. 


\begin{figure*}[t]
    \centering
    \includegraphics[scale=0.22]{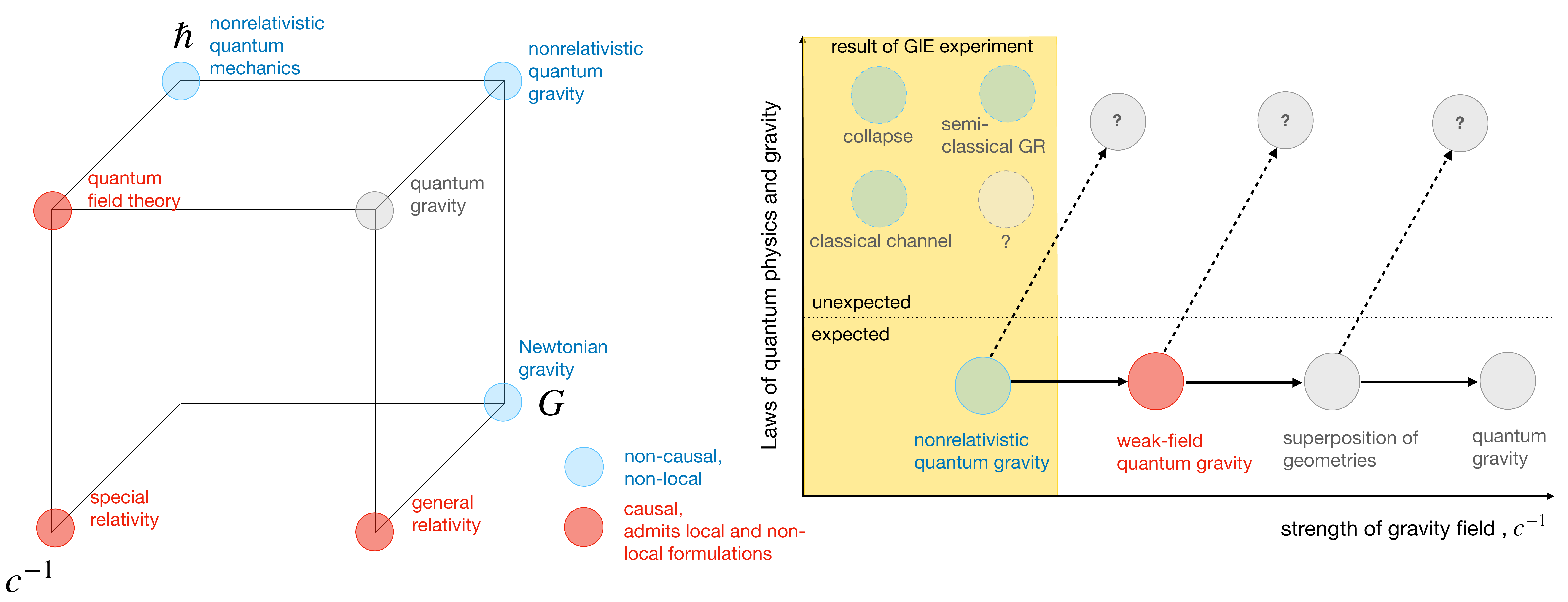}
    \caption{\small \textit{Left:} Bronstein cube\cite{Bronstein1933, okun2002cube} organising various theories depending on their dependence on $\hbar, G$ and $c$. The GIE experiment tests ``non-relativistic quantum gravity'', which is a non-local and non-causal theory (due to Newtonian gravity being instantaneous action-at-a-distance). Since in the $\hbar \rightarrow 0$ limit General relativity is  geometric, causal and admits a local formulation, it is plausible that quantum gravity itself should share these features -- but a complete theory is unknown as indicated by a grey circle. We note that quantum field theory and special relativity admit non-local causal formulations, such that it is plausible that also quantum gravity might admit a non-local formulation. Therefore entanglement may also be generated non-locally in the fully relativistic theory, and LOCC does not imply quantization of mediators.
        \textit{Right:} The yellow region indicates the direct experimental access of the GIE where gravitational fields are weak and matter non-relativistic. An unexpected result (dashed circles) would rule out non-relativistic quantum gravity.
    For an expected result, at the bottom, we outline the ``standard'' view how non-relativistic quantum gravity is connected (black arrows) to quantum gravity. 
    However, this is merely a reasonable extrapolation and it is conceivable that the fundamental theory deviates from these properties (grey circles with question marks that may have conventional theories as limiting cases, indicated by dashed arrows). Thus observing the expected outcome of the GIE experiment only directly confirms non-relativistic quantum gravity and everything else is an extrapolation based on ones' theoretical priors.   }
    \label{fig:setup2}
\end{figure*}

\section{The gravitational case: quantized mediators or non-local alternatives?} \label{sec-GR}
Above we discussed the electromagnetic version of the GIE experiment. We showed that conclusions beyond the non-relativistic limit are ambiguous because there is always an alternative explanation possible that generates entanglement non-locally, thus the LOCC argument does not unambiguously imply quantization of mediators. It is conceivable that similar dual arguments based on either a local mediator theory or a non-local theory could apply to quantum gravity. This is indeed the case for weak gravity.

\subsection{Local mediators: Weak-field quantum gravity in Lorentz gauge} \label{sec:LorentzWeakfieldgravity}

In the weak-field limit, covariant quantization of gravity has been developed in Refs.\cite{1952PPSA...65..161G,PhysRev.172.1303} by making use of the so-called indefinite metric. This procedure is very similar to the covariant quantization discussed previously for the case of QED. In this formulation, all the components of the (linearized)  metric tensor $h_{\mu\nu}$ are quantized. The fictitious-gauge degrees of freedom are eliminated by imposing supplementary conditions on the Hilbert space.  Classically the Lorentz gauge (or de Donder gauge) is defined as   $\partial^\mu \bar h_{\mu\nu}=0 $ with  $\bar h_{\mu\nu}\equiv h_{\mu\nu} - 1/2\, h_{\mu}^\mu$.  The procedure in brief is the following: We expand the gravitational field in Fourier basis,
\begin{multline} \label{weakfield_mode_decomposition}
    \hat{h}_{\mu\nu}(\vec{r})=\int d^3k\sqrt{\frac{ \hbar G}{\pi^2\omega_k}} \biggl[\hat{a}_{\mu\nu}(\vec{k})e^{i\vec{k}\cdot\vec{r}}+\\+\hat{a}_{\mu\nu}^{\dagger}(\vec{k})e^{-i\vec{k}\cdot\vec{r}}\biggr]\,.
\end{multline}
The Hamiltonian of the gravitational field has the form \cite{1952PPSA...65..161G}
\begin{equation}
    \hat{H}_g=\frac{1}{2}\int d^3k \hbar\omega_k \biggl[\hat{a}^{\dagger}_{\mu\nu}(\vec{k})\hat{a}^{\mu\nu}(\vec{k})-\frac{1}{2}\hat{a}^{\dagger\mu}_{\mu}(\vec{k})\hat{a}^{\nu}_{\nu}(\vec{k})\biggr]
\end{equation}
where the graviton creation and annihilation operators satisfy the commutation relations 
\begin{multline} \label{weakfield_commutation_relation}
    [\hat{a}_{\mu\nu}(\vec{k}),\hat{a}^{\dagger}_{\lambda\rho}(\vec{k}')]=\bigl(\eta_{\mu\lambda}\eta_{\nu\rho}+\eta_{\mu\rho}\eta_{\nu\lambda} \\ -\eta_{\mu\nu}\eta_{\lambda\rho}\bigr)\delta(\vec{k}-\vec{k}').
\end{multline}
where $\eta_{\mu\nu}$ is the Minkowski metric. Note that the $\hat{a}_{0i}(\vec{k})$ for $i=1,2,3$ components as well as the trace $\hat{a}^{\mu}_{\mu}(\vec{k})\equiv\hat{a}(\vec{k})$ satisfy the commutation relations with a negative sign: 
\begin{equation}
    [\hat{a}_{0i}(\vec{k}),\hat{a}^{\dagger}_{0i}(\vec{k}')]=[\hat{a}(\vec{k}),\hat{a}^{\dagger}(\vec{k}')]=-\delta(\vec{k}-\vec{k}')
\end{equation}
thus an indefinite Hilbert space metric has to be used.\cite{1952PPSA...65..161G,PhysRev.172.1303} Matching up the components of this metric which have negative commutation relations along with components of the metric with positive commutation relation, one can eliminate the eight redundant graviton polarizations to end up with the two physical degrees of freedom. 

In presence of interactions, the supplementary condition needs to take into account the presence of matter degrees of freedom \cite{1952PPSA...65..161G}
\begin{equation}
\begin{split}
\biggl[&\hat{a}_{3\nu}(\vec{k})-\hat{a}_{0\nu}(\vec{k})
    + \\
 & + \kappa\int_{t=t'}d^3x' D^{(+)}(\vec{x}-\vec{x}')\hat T_{0\nu}(\vec{x}')\biggr]|\chi\rangle=0
\end{split}
\end{equation}
for $\nu=0,1,2,3$ where $D^{(+)}(\vec{x}-\vec{x}')$ is the positive frequency part of the graviton propagator and $\hat{T}_{0\nu}(\vec{x}')$ corresponds to the matter stress energy tensor. In this case, the redundant gravitons are still absent in all physical processes, but they can exist in virtual states. In connection to GIE, this quantization procedure has been adopted in Ref.\cite{bose2022mechanism} in order to argue that the masses get entangled via the exchange of virtual gravitons.

\subsection{No mediators: Weak-field quantum gravity in Poisson gauge} \label{sec:PoissongaugeWeakfield}

Up to this point, we have discussed the covariant quantization of  weak-field gravity. In this picture, the virtual gravitons that correspond to the many components of the metric tensor, even though they are not the usual physical gravitons and are unobservable, exist in the mathematical formulation of the theory. It is possible however, to formulate the theory in a picture where no redundant and unphysical gravitons appear at any stage in the theory. In this formulation, one eliminates the redundant degrees of freedom completely by fixing the gauge classically, and then canonically quantizes the remaining physical degrees of freedom \changes{$\hat s_{ij}^{TT}$ which are the spatial and transverse traceless components of $\hat h_{\mu \nu}$ (as defined in eq.\,\eqref{weakfield_mode_decomposition}). See App.\,\ref{sec:Appendix_weak_field_poisson} for further details on the isolation of the physical degrees of freedom}.  The Hamiltonian in this so-called Poisson gauge -- analogous to the Coulomb gauge in QED -- is given by  
\begin{align} 
\hat{H}_{int}=& -\int d^3r\bigl(\hat s_{ij}^{TT}(\vec{r}) \hat T^{ij}(\vec{r})\bigr)- \label{hamiltpoisson}\\
&
-\frac{G}{2}\int\frac{d^3rd^{3}r'}{|\vec{r}-\vec{r}'|}\biggl[ -4\hat f_{\perp,i}(\vec{r}')\hat T^{0i}(\vec{r})  + \notag\\
\quad &  +\hat T_{00}(\vec{r}')\bigl(\hat T^{00}(\vec{r})+\hat T^{k}_{k}(\vec{r})-2\hat \Pi^{||}(\vec{r}) \bigr) \biggr] \,.\notag
\end{align}
\changes{All terms except the first one are} 
spatially non-local functions of the stress-energy tensor. \changes{Note that $\hat f_{\perp,i}(\vec{r})$ and $\hat \Pi^{||}(\vec{r})$ are defined via the stress energy tensor}. For further details and derivation of \eqref{hamiltpoisson}, the reader is referred to Appendix  \ref{sec:Appendix_weak_field_poisson}.

The Hamiltonian in Poisson gauge \eqref{hamiltpoisson} can generate entanglement between two masses through the exchange of the physical spin-2 gravitons (the term $\hat s^{TT}_{ij}$) and/or the non-local interaction between the quantized matter degrees of freedom (2nd and 3rd line, which are both non-local). In the regime in which gravitational radiation is irrelevant, as it is the case for the GIE proposal, the entanglement is generated non-locally via direct matter interactions. 
In direct analogy to the toy model for QED entanglement generation, as discussed in section \ref{sec-EM}, one can now directly compute how the two masses entangle in this non-local formulation. The computation carries over one-to-one: 
the non-local term in \eqref{hamiltpoisson} that involves
$-\frac{G}{2}\int\frac{d^3rd^{3}r'}{|\vec{r}-\vec{r}'|}\hat{T}_{00}(\vec{r}')\hat {T}^{00}(\vec{r})$ generates the entanglement, since all other terms are negligible. No mediators appear at any stage of the calculation. In contrast, the authors of Ref.\cite{bose2022mechanism} arrive at this same interaction term above in the local Lorentz gauge, where quantized virtual mediators appear as discussed in the previous section -- the two physical predictions are thus equivalent, as expected, but the ontology can be local or non-local. Additionally, also in the Lorentz gauge the authors show explicitly that no virtual spin-2 gravitons appear in the non-relativistic limit relevant for GIE, only the auxiliary, unobservable mediators. Thus neither in the Lorentz gauge nor in the Poisson gauge do the spin-2 gravitons contribute (their contribution is gauge invariant), not even in virtual processes. 

Overall, from the non-local picture presented here in the Poisson gauge, the relevant interaction term appears without the involvement of any local mediators. Thus one cannot use the LOCC argument to infer the existence of quantized mediators in the gravitational case either.

\changes{In this context, we also refer to Zel'dovich and Novikov \cite{ZeldovichNovikov1971}, who say: 
\emph{[a] popular but untrue assertion is that the gravitational interaction is due
to an ``exchange of gravitons.'' [...] The difference in character between the longitudinal and the transverse 
fields clearly illustrates the absurdity of a literal interpretation of the phrase, 
``Interaction is an exchange of quanta.''}
While this quote highlights the caveats of a strictly local interpretation and emphasizes the equivalent non-local one, in our view the two are simply equally viable. Any preference boils down to imposing additional assumptions about the nature of physics that are not distinguishable in relativity, such as imposing that only physical degrees of freedom are quantised, or alternatively that physics is strictly local. Only if the latter is imposed, such that the equivalent non-local description is simply excluded by hand, can the LOCC argument be used to indirectly infer quantized mediators of gravity.}


\subsection{No mediators: Weak-field quantum gravity as absorber theory} \label{sec:grav-absorber}
Additionally, in analogy to electromagnetism, in the weak-field regime of gravity there also exists an absorber formulation,\cite{Rosen1979, Louis-Martinez2012} which would comprise a weak-field quantum gravity after source mass quantisation. Thus the same ambiguity that we highlighted in the electromagnetic case also exists in the weak-field limit of General Relativity: Mediators can be fully removed from the theory by changing to a fundamentally non-local source and absorber ontology. Within this framework alone, it is therefore not possible to conclude that entanglement is generated through quantized, local mediators from the outcome of the GIE experiment and validity of Post-Newtonian gravity alone.

However, no absorber formulation of full general relativity has been found so far. 
It is currently not known if a classical absorber theory for gravity can exist at all.
Attempts to construct absorber theories of non-Abelian gauge theories and gravity are incomplete,\cite{HoyleNarlikar1972, BurgersVanDam1987, Turygin1986, Vladimirov2008, Louis-Martinez2012Fields, Louis-Martinez2012} or failed.\cite{GibbonsWill2006} Note that the gravitational theory developed by Hoyle and Narlikar \cite{HoyleNarlikar1964} is not an absorber theory of gravity since the metric appears as degree of freedom in the action.\cite{DeserPirani1965, HoyleNarlikar1966}
There are gravitational absorber theories that correctly reproduce the Post-Newtonian limit of GR for N particles, the Einstein-Infeld-Hoffman equations,\cite{EinsteinInfeldHoffmann1938} see Refs.\cite{Louis-Martinez2012,BolokhovKlenitsky2013}.
This limit however does not include the effects of gravitational radiation.
Thus, whether a loophole to the LOCC argument in form of a non-local absorber theory applies also to gravitational interactions depends on whether one can construct a complete absorber theory for gravity, which is a fascinating open question.  \cite{WesleyWheeler2003}

It is nevertheless conceivable that any hypothetical quantum theory of gravity that is theoretically acceptable and experimentally viable could share the ambiguity of QED of whether entanglement is produced through virtual longitudinal and scalar gravitons, or through a direct non-local interaction of quantized source masses.
In the absence of such a theory of quantum gravity, the conceivable possibilities for interpreting gravitationally induced entanglement can only be broader. 
There are even arguments that suggest that any quantum gravity might exhibit fundamental non-locality. \cite{GiddingsLippert2004, GiddingsMarolfHartle2006, Giddings2006} 
Thus, observing gravitationally induced entanglement does not tell us whether gravity is quantized unless we insist on assuming that quantum gravity is a local quantum field theory  -- and only the local ontology is permissible.
The resilience of gravity to quantization and certain properties of gravity that suggest that it would have a locally finite dimensional Hilbert space, could be an indication that quantum gravity is at odds with being fundamentally a local QFT. We thus have two possible sets of assumptions for general relativity: either being fundamentally a local quantum field theory or having some non-local interpretation. Neither our theoretical understanding today nor the GIE experiment can distinguish these two. Thus also for gravity the conclusion from the GIE experiment about quantized mediators depends on what prior assumptions we are willing to accept, but that are not imposed on us from our current understanding of physics.

\section{Path integrals} \label{sec-path}
\subsection{Electromagnetic case}
To further clarify the differences in the various interpretations, here we consider in the electromagnetic case the path integrals involved in the non-relativistic Coulomb case, Coulomb gauge QED, Lorenz gauge QED, and absorber theory. This formulation helps highlight the apparent paradoxes and their resolutions and fully clarifies the relation between all 4 cases in the electromagnetic case. Interestingly, the mathematical starting points of the various QED formulations are very different, but for the case of QED they lead to identical physical outcomes.

The path integral approach to QED assumes that any amplitude between quantum states of particle configurations and the electromagnetic field can be calculated from the path integral
\begin{equation}
  \int \mathcal{D} \mathcal{F}' \mathcal{D} x' \exp\left( \frac{i S}{\hbar} \right)  
\end{equation}
with the action $S=S[x'_a(t), \mathcal{F}'(x,t)]$ being a functional of the particle trajectories $x'_a(t)$,  where $a=1,2$ labels the two particles, and the field configuration ``trajectories'' $\mathcal{F}'(x,t)$. With the measures $\mathcal{D} \mathcal{F}' $ and $\mathcal{D} x'= \Pi_a \mathcal{D} x'_a$ the path integral sums independently over all conceivable field configuration trajectories $\mathcal{F}'$ and particle trajectories $x'_a(t)$.

Since the electromagnetic field is a gauge field with unphysical degrees of freedom care needs to be taken when performing the path integral $\mathcal{D} \mathcal{F}'$. There are several options. One can fix the gauge completely before performing the path integral, such as in Coulomb gauge, or one can include a gauge fixing term in the path integral. The latter can be effectively achieved by quantizing in Lorenz gauge following Gupta and Bleuler.\cite{Gupta:1949rh,Bleuler:1950cy}

The theory is called local if the Lagrangian density $\mathcal{L}[x'_a(t), \mathcal{F}'(x,t)](x,t)$ consists only of local polynomials of the current $J[x'_a(t)](x,t)$ and fields $\mathcal{F}'(x,t)$ (and its first and second derivatives). An example for a local Lagrangian is $\mathcal{L}(x,t)=J[x'_a(t)](x,t) \mathcal{F}'(x,t)$, whereas an example for a non-local Lagrangian density is $\int d^4 x' J(x',t') D(x'-x,t'-t) J(x,t) $ with some kernel $D$. In any case the action is given by $S = \int d^4 x \mathcal{L}(x,t)$.

\subsubsection{QED in Lorenz gauge} 
In the Lorenz gauge the path integral involves the two  physical (transverse) photon polarizations and two unphysical auxiliary photons, the so-called scalar and longitudinal photon, see Ref.\cite{Cohen-TannoudjiEtal1992}, so that 
\begin{equation}
  \int \mathcal{D} \mathcal{F}_{\rm phys}' \mathcal{D}\mathcal{F}_{\rm aux}' \mathcal{D} x' \exp\left( \frac{i S}{\hbar} \right)
\end{equation}
where $S=S[x'_a(t), \mathcal{F}'_{\rm phys}(x,t),\mathcal{F}'_{\rm aux}(x,t)]$.
Keeping these auxiliary photons in the path integral allows the Lagrangian to remain manifestly Lorentz invariant and local.

\subsubsection{QED in Coulomb gauge} If we chose to integrate only over physical degrees of freedom, which can be conveniently done by fixing the gauge completely to Coulomb gauge, we obtain the  path integral
\begin{equation}
  \int \mathcal{D} \mathcal{F}_{\rm phys}'  \mathcal{D} x' \exp\left( \frac{i S}{\hbar} \right)
\end{equation}
where  $S=S[x'_a(t), \mathcal{F}'_{\rm phys}(x,t),\mathcal{F}_{\rm aux}[x'_a(t)]]$ and with $\mathcal{F}_{\rm aux}[x'_a(t)]$ denoting the on-shell, or classical, solution of the auxiliary field sourced by the charges at $x'_a(t)$. The Lagrangian density lost its manifest Lorentz invariance and its locality. More importantly since $\mathcal{F}_{\rm aux}[x'_a(t)]$ is instantaneously determined by the charge distribution $x'_a(t)$ at the same time, physics appears to have become acausal in Coulomb gauge.
But this is not the case as discussed further below.

\subsubsection{QED absorber formulation}
In the absorber theory the path integral only involves the charges:
\begin{equation}
  \int  \mathcal{D} x' \exp\left( \frac{i S}{\hbar} \right)
\end{equation}
with  $S=S[x'_a(t), \mathcal{F}[x'_a(t_{\rm ret})]]$. Note in contrast to \eqref{NonrelPathintegral} where the dependence of $\mathcal{F}$ on the trajectories is  instantaneous it is here at the retarded time and thus causal, see Ref.\cite{Davies1972}.
Note that we have here omitted the half-advanced-half-retarded contribution since radiation reaction effects can be neglected in the GIE experiment.
Relevant to the GIE experiment is the case of weak-field gravity where an equivalent absorber action can be easily constructed which has been used in \cite{ChristodoulouEtal2022} to simplify calculations of the GIE. Equation 9 in Ref.\cite{ChristodoulouEtal2022} is the same as the absorber action with the radiation reaction contribution neglected, so that a fully retarded action and the absorber action lead to indistinguishable results. The crucial point is that the expression only involves the charges, and not the fields. While this was an approximation in Ref.\cite{ChristodoulouEtal2022}, we argue here that it could instead be used as the starting point of the whole analysis.

\subsubsection{Non-relativistic Coulomb interactions}
The analog of Newtonian gravity is Coulomb theory which is electrodynamics with everything but the scalar photon discarded. This leaves us with 
\begin{equation} \label{NonrelPathintegral}
  \int  \mathcal{D} x' \exp\left( \frac{i S}{\hbar} \right)
\end{equation}
where  $S=S[x'_a(t), \mathcal{F}_{\rm aux}[x'_a(t)]]$ and $\mathcal{F}_{\rm aux}[x'_a(t)]$ is the solution to the Coulomb equation given the charge distribution of $x'_a(t)$. Like in the Coulomb gauge of QED  $\mathcal{F}_{\rm aux}[x'_a(t)]$ is instantaneously determined by the $x'_a(t)$. As we will discuss further below, the absence of $\mathcal{F}'_{\rm phys}(x,t)$ is responsible for the physics to be genuinely acausal in this theory.

\subsubsection{Equivalent predictions}
Interestingly, the path integrals for QED in Lorenz gauge, QED in Coulomb gauge and the absorber theory, all lead to identical predictions for amplitudes after renormalization. \cite{HoyleNarlikar1995} It thus becomes obvious that the question of ``what produces the entanglement in the electromagnetic GIE experiment?'' is ambiguous, even in the relativistic descriptions. 
To answer the question in all these cases we just need to look at what determines the dominant part of phase factors of the four different semi-classical paths of the two particles (See Figure 1, where the four distances are LL, LR, RL, RR).
In Lorenz gauge QED the phase is due to the auxiliary fields a.k.a ``virtual photons'', in  Coulomb gauge QED and the Coulomb theory it is due to auxiliary fields slaved to the particles  due to the constraint equation, and in the absorber theory it is due to direct particle interactions on the lightcone in the complete absence of any field. Coulomb gauge QED, non-relativistic Coulomb theory and absorber theory thus share the property that entanglement is established without any mediators. 

\subsubsection{Causality}

The path integral formulation again suggests that apart from the picture in the Lorentz gauge, entanglement in QED seems to be established in the same acausal manner as in Coulomb theory, even though QED is a fundamentally causal theory. 
This arises because in QED the physical photons  do not contribute to the entanglement in the proposed experiment in the limit of very slow wave-packet splitting and merging, and thus the entangling phase factor in the path integral is dominated by the Coulomb field that is sourced instantaneously from the charge distribution rather than the retarded charge distribution. 
Nevertheless the QED results are fully causal in all formulations: the on-shell $\mathcal{F}'_{\rm phys}[x_a](x',t)$ combined with $\mathcal{F}_{\rm aux}[x'_a(t)](x,t)$ leads to a manifestly causal expression, which coincides with the expression appearing in the absorber theory. Interestingly, this precise expression has been explicitly used in Ref. \cite{ChristodoulouEtal2022} (not in the context of absorber theories) to establish that no entanglement would be produced in a GIE type experiment if the splitting and merging of the wave packets happened at spacelike separations. We thus see that while $\mathcal{F}_{\rm aux}$ dominates the phase factor in the GIE experiment, one needs to include $\mathcal{F}_{\rm phys}$ to restore causality, or generally whenever a near field approximation is not a good approximation and retardation effects become important. At spacelike distances the would-be entanglement of $\mathcal{F}_{\rm aux}$ is exactly cancelled by $\mathcal{F}_{\rm phys}$. 

\subsection{Gravitational path integrals}

The discussion carries over analogously to linearized gravity. In the case of nonlinear gravity, the equivalence between a field path integral and an absorber theory will likely break down. However one could simply postulate a theory of quantum gravity in which the path integral only involves matter and the gravitational field is ``slaved'' to the matter configuration like in an absorber theory. In other words, one would associate with each quantum state of matter a correspondingly sourced gravitational field and allow for coherent superpositions of such source-gravity entangled states. This type of ``simplified'' quantum gravity model can be a useful approach, as in e.g. Refs. \cite{zych2019bell, giacomini2020einstein} and
 would correctly predict the GIE experiment because it automatically has the same linearized weak-field limit that was assumed to hold for a ``proper'' quantum gravity theory in which geometries are summed over in the path integral independently of the matter.\cite{ChristodoulouEtal2022} 
The two situations would likely only be distinguishable
in the strong field regime where graviton loops contribute in the latter case but not in the former. As long a we limit ourselves to weak gravity, however, the equivalence between the local and non-local path integral formulations as discussed in the QED case remains.

\section{Cosmology and signatures of quantum gravity} \label{sec:cosmo}
In the previous chapters we have considered the GIE experiment from the point of view of different  theoretical (or even meta-physical) assumptions. Here we discuss how the same assumptions applied to cosmology can provide similar conclusions from already existing observations. 

Quantizing  the weak-field linearized gravitational field is standard procedure within inflationary cosmology.\cite{MukhanovChibisov1981, MukhanovChibisov1982, Mukhanov2005}
In this scenario the Universe's energy density is initially dominated by a slowly evolving scalar field $\varphi$ in an approximately homogeneous and isotropic state $\bar{\varphi}$, which according to General Relativity leads to an approximate de Sitter state of spacetime in which vacuum fluctuations of the metric and the scalar field are amplified and stretched to superhorizon scales. These fluctuations then seed the cosmic large scale structure after the end of inflation by giving rise to local variations in the energy density of radiation. This in turn is perceived as temperature anisotropies in the Cosmic Microwave Background today.
The full Einstein-Hilbert and scalar field action is expanded around the time-dependent Friedmann-Robertson-Walker background with scale factor $a(t)$ so that the scalar fluctuations $v$ are governed by the action
\begin{equation}
    S = \frac{1}{2} \int d^4 x (v'^2 - \partial_i v \partial_i v + \frac{z''}{z} v^2)\,.
\end{equation}
Here $'$ denotes a derivative with respect to conformal time, $\mathcal{H}=a'/a$ is the conformal Hubble parameter and 
 $v(t,\vec x) = a(\delta \varphi(t,\vec x) + \psi(t, \vec x) \bar \varphi'/\mathcal{H})$ is a combination of the scalar field fluctuation  $\delta \varphi = \varphi - \bar \varphi$ and the Newtonian scalar perturbation of the spatial metric $g_{ij} = a^2 (1- 2 \psi) \delta_{ij}$.\cite{Mukhanov1988} The function $z= a \bar \varphi'/\mathcal{H}$ is purely time dependent, and thus classical. The quantity $v(t,\vec x)$, however, is quantized.
The gauge invariant\footnote{Note that while a gauge transformation could remove the metric fluctuation $\psi$ from the variable $v$, quantisation of $v$ still implies the quantisation of scalar metric perturbations via the perturbed Einstein equation \eqref{NewtonianGaugeInflaton}.} quantized gravitational field $\hat \Phi(t,\vec x)$ is then given by
\begin{equation} \label{NewtonianGaugeInflaton}
    \nabla^2  \hat \Phi(t,\vec x) = 4 \pi G \frac{\bar \varphi'{}^2}{\mathcal{H}^2} \Big(\frac{\hat v(t,\vec x)}{z}\Big)' .
\end{equation}
This is the quantized Newtonian gravitational potential as we discussed in section \ref{sec-NR}, sourced by the quantized perturbations of the inflaton field.
Only if the gravitational field is quantized in this way, and in essentially\footnote{The only difference is that the weak-field is quantized around a Friedmann-Robertson-Walker background instead of a Minkowski background and instead of particles the source is a scalar field. Apart from this, equations \eqref{NewtonianGaugeInflaton} and \eqref{NewtonQuantised} have the same physical meaning.} the same way as required for the GIE experiment to produce the expected entanglement, does the inflationary paradigm predict metric fluctuations with variance $\langle 0 | \hat \Phi(t,\vec x) \hat \Phi(t,\vec x+ \vec r) | 0 \rangle $ with an amplitude and $r$-dependence in accordance with that observed in cosmological observations.\cite{PlanckInflation2020} \changes{This variance is directly linked to the observed angular dependence of the temperature anisotropy variance of the cosmic microwave background. More specifically, on large scales the observed fluctuation in temperature $ \hat T(t_0, \vec{x}_0, \vec n) /\bar T(t_0) -1$ in a given direction $\vec n$ is proportional to the Newtonian potential $\hat \Phi(t_{r}, \vec{x}_0 -  \vec{n} t_0)$ on the corresponding location at the last scattering surface, when photons last scattered at time of recombination $t_{r}$. This is predominantly due to the gravitational red-shift, known as Sachs-Wolfe effect, of photons climbing out of gravitational potential wells}.\cite{Mukhanov2005} The quantum nature of the sources for the gravitational Newtonian field are essential, as their quantum fluctuations (uncertainty) produce the observed signal.
Thus, the observed angular dependence of temperature fluctuations -- and some of its qualitative features on the largest scales, 
such as anti-correlations of temperature and polarisation fluctuations on super horizon scales, as well as the small increase of power towards the largest scales (``red tilt'') -- can be considered evidence for the necessity of quantized gravity.

The prediction of the CMB temperature fluctuation correlation from inflation thus required the quantisation of the weak-field gravitational field many decades before the GIE proposal. 
Several generic a priori predictions based on canonical single field inflation made 40 years ago \cite{MukhanovChibisov1981, MukhanovChibisov1982} have been confirmed observationally, such that one might say that evidence for quantizing weak-field gravity already exist today. 
Indeed, mean-field theory as mentioned in Sec.\,\ref{sec-NR} would not be adequate to predict the temperature fluctuations of the Cosmic Microwave Background within the inflationary paradigm, as no fluctuations  would be produced at all during inflation.

In principle the inflationary paradigm also predicts the amplification and stretching of gravitational radiation which leaves a distinctive imprint in the polarisation of the temperature fluctuations, so-called primordial B-mode polarisation. Currently there are only upper bounds on the amplitude of this type of anisotropy which has already ruled out certain simple shapes of the inflaton potential.\cite{PlanckInflation2020}  Observation of primordial B-modes would be evidence for the quantum nature of the spin-2 graviton. But as discussed above, within the assumptions of the GIE proposal, already the CMB fluctuations as observed indicate the quantisation of the Newtonian part in similar spirit to the GIE proposal. 

Finally, we note that fields evolving under their self-gravity or on an expanding space can exhibit more intricate quantum features.
For instance, the quantum states of the inflaton or of a light scalar field comprising the dark matter, as well as the quantum state of gravitational radiation can get squeezed.\cite{GrishchukSidorov1990, AlbrechtFerreiraProkopecEtal1994, KoppFragkosPikovskiPRD, KussMarsh2021} Cosmological observations might thus provide unique tests of weak-field quantum gravity,\cite{KannoSoda2019, Martin2019, GreenPorto2020} or even of genuine superpositions in more speculative scenarios, \cite{pikovski2016quantum} should these scalars fields exist.

\section{Discussion} \label{sec-discussion}

\subsection{Newtonian quantization in GIE and cosmology }
What then does the GIE experiment teach us in addition to current cosmological observations, as discussed above? First of all, not everyone might believe in the inflationary paradigm: the conclusion that gravity is quantized requires a prior belief in the current standard inflation paradigm where it is sourced by quantum fields. Furthermore, while cosmological observations of the cosmic microwave background are very accurate, the object that is used to study the quantisation of gravity has been created more than 13 billion years ago and modulated by unknown components such as dark matter and dark energy. Any independent and more direct evidence for the quantisation of weak-field gravity is thus still very valuable.
Furthermore, the GIE experiment probes more directly the fact that the gravitational field can be put into superposition, which is necessary for entanglement between source masses to arise.
 In contrast, in the cosmological setting it is quantum fluctuations of masses that source fluctuations in the gravitational field. 
While this is a direct consequence of quantizing the gravitational field, it is conceivable that this effect can be mimicked by an entirely classical theory, e.g. the standard classical weak-field Einstein equations augmented by a stochastic yet classical noise term that sources fluctuations maybe during a (classical) inflation. In other words it is conceivable that the gravitational field is never put into superposition or any quantum state, albeit considered implausible in the cosmology community.
The strength of the GIE experiment lies in the difficulty of alternative explanations for producing gravitationally induced entanglement that would not require the gravitational field to exist in superposition during the experiment -- there are however some suggestions for alternative models \cite{HallReginatto2018, DoenerGrossardt2022} in which LOCC does not necessarily hold since quantum physics is modified.
Overall, observing gravitationally induced entanglement would give more direct support to the idea that the gravitational field can exist in superposition.

\subsection{Ambiguous and unambiguous conclusions from the GIE experiment}

As we have shown, the conclusions from GIE vary depending on what set of assumptions one uses as a starting point. At the core, the experiment probes what has been discussed in section II: 
The expected result is based on non-relativistic quantum mechanics with gravitational interactions, that is quantized source masses interacting  through a Newtonian potential $ \hat \Phi= -G m_1 m_2/|\hat{\vec{x}}_1-\hat{\vec{x}}_2|$. Gravity is introduced into the non-relativistic quantum theory  the same way as the electric Coulomb potential. The experiment, if successful,  thus unambiguously confirms that gravity is sourced coherently by a mass in superposition, from which one can conclude that Newtonian gravity is coherently associated with a spatial superposition of the source. To date such a superposition of Newtonian potentials remains unverified and untested, thus GIE would provide a test of physics in a new regime. The experiment can also rule out alternative models, such as a mean-field source for the gravitational potential or a classical channel model.\cite{kafri2014classical} And of course any unexpected result would hint at new physics.

Going beyond these conclusions, however, is ambiguous. The original aim of the GIE proposals was to draw conclusions about the nature of quantum gravity by assuming that this is a relativistic and local theory of which the  non-relativistic limit with its inherent non-locality is merely an artefact. But as we have shown, even if one accepts this reasonable starting point, still no unambiguous conclusion about the underlying mediators can be made. This is because non-local entanglement generation is perfectly allowed even within relativistic physics. 
QED is a good example that locality is not a necessary assumption and does not need to be imposed on such a relativistic parent theory -- as opposed to causality. This is not simply an issue of gauge, but that locality is not necessarily satisfied for the underlying ontology even in relativity. This is important here, as the proposed tests are of indirect nature and the underlying ontology is central for the interpretation of the outcomes. If a non-local explanation is possible, then LOCC is simply not able to distinguish between a quantum and a classical mediator. Thus, while the absorber formulation may be considered as a ``quirk'' of the theory, it still shows that locality needn't be satisfied at the fundamental level, even if all observations are causal. For gravity, it remains an open question if an absorber theory can capture all of general relativity, but it exists for the weak-field limit. Moreover, it is conceivable that quantum gravity is fundamentally non-local in some way. In all these cases the LOCC argument breaks down, since the locality condition is no longer satisfied and thus observation of entanglement generation does not necessarily imply quantised gravitational mediators.

Thus it is only when locality is imposed as an \textit{additional} condition that one can conclude quantization of mediators in such indirect tests. While we agree that the additional assumption of locality is not just allowed but also plausible and intuitive,\cite{vedral2020local,marletto2021interference}  we have no indication that it must be satisfied in nature at the fundamental level. 
A serious drawback is that imposing locality introduces additional, fictitious degrees of freedom such as scalar and longitudinal photons in QED, and analogous types of new gravitons in weak-field quantum gravity, that can never be directly observed. Yet they are responsible for mediating entanglement in this picture. The other picture, which is non-local, gets rid of these unphysical degrees of freedom, and thus has its own merits. Fundamentally, causality seems to be the underlying principle in all these formulations, and locality only a convenient consequence in one of them. Thus, the only unambiguous statement one can draw from the GIE experiments with regards to mediators is that if one pre-assumes an underlying local field theory, then such a field theory would need to include quantized mediators. But this assumption is very restrictive in that it artificially excludes perfectly valid non-local formulations. Our cosmology example shows that if the same assumptions about the nature of gravity are accepted, in addition to the standard inflationary scenario, then already the observed CMB fluctuations are sufficient to deduce quantum fluctuations of the gravitational field. 

\subsection{Agents and locality}
From a quantum information perspective the GIE proposal involves two localised agents (Alice and Bob) situated close to their respective particle. Each agent directly manipulates only their respective nearby particle locally (through the particle's spin in the GIE proposal\cite{BoseAnupam}) so as to create the spatial superposition and subsequently  interfere the particle's wave function to infer the creation of entanglement. If, however, the two particles interact non-locally, then  Alice's  local interaction with her particle will inadvertently and unavoidably also affect Bob's particle non-locally, and vice versa. Thus the GIE proposal's experimental setup does not guarantee that the two agents perform only local operations because the gravitational interaction between the particles is not necessarily local.  The LOCC theorem in combination with witnessing entanglement does therefore not imply the existence of quantized mediators since a necessary condition for the theorem to apply -- locality -- is not necessarily present and cannot be tested for.

As we discussed in Sec.\ref{sec:nomedQEDabsorber}, in the specific non-local theories studied in this paper in which quantum physics is unmodified, entanglement is necessarily and by construction produced quantum mechanically.
Our point is that the existence of the quantum nature of the underlying system that generates entanglement  could not have been inferred by merely invoking the LOCC argument alone. 
Even if entanglement is observed, it may be produced by some process that may correspond to some underlying system that is not quantum mechanical in the usual sense.\cite{Popescu1994-POPQNA,Pal2021experimental} We note here that it is therefore also interesting to exactly quantify the expected entanglement, and probe for possible deviations.


\subsection{Superpositions of geometries}
\label{subsec:superposed_geometries}
Can one draw conclusions about the quantum nature of gravity beyond mediators?  It was pointed out in several previous works \cite{RovelliChristod2019, ChristodoulouEtal2022} that it is not the quantum nature of mediators the GIE proposal is sensitive to, but the ability of the gravitational field to exist in a quantum superposition.
Can one gain insights on quantum gravity from that alone?

A compelling interpretation is that the successful experiment shows that geometries of space-times exist in superposition. This is based on additional assumptions, that we have not yet discussed: The extrapolation of classical general relativity (GR) and its ontology into the quantum domain. In this view, the Newtonian limit is just a limiting case of what is space-time geometry, and any superpositions of the Newtonian field correspond to superpositions of geometries in the context of GR. One way to see this is in the path integral formulation of the gravitational case: 
The phase difference between the semi-classical paths of the particles is dominated by solutions of the classical Einstein equations sourced by the semi-classical matter configurations entering the superposition.\cite{ChristodoulouEtal2022}
In this sense the GIE experiment would show that geometry can be put in superposition. 
This is true no matter which gauge one adopts, or if one thinks of the geometry as fundamental or arising from virtual gravitons, the constraint equation, or direct particle interaction on the light cone. In all cases superposition of effective geometry is the origin of the entanglement. 
The only way to avoid this conclusion is to find a viable gravitational theory that is not geometrical in the quantum regime and/or to replace the path integral, and thus quantum mechanics, by some other theory in this limit.
While we are not aware of any theory of gravity consistent with observations that is not geometrical, it is conceivable that the path integral formula is not correct if contributions of different geometries would lead to an observable effect. For instance in Penrose's gravitational collapse mechanism, if the wave packets in the GIE experiment were kept a sufficiently long time apart before they are attempted to be merged, only one of the four semi-classical contributions to the path integral (LL, LR, RL, RR)  would survive according to Penrose (the other branches are ``dying'' before the matter wave packets are attempted to be merged, and thus no merging actually happens), in which case we might expect the complete absence of entanglement and any other interference effect of the particles. In general, a modification of the path integral that would lead to the expected entanglement without having geometries in superposition would have to be drastically different from standard quantum mechanics. For instance we could imagine that each path was weighted with a phase produced by the gravitational mean field (thus geometry is not in a superposition) but that there is an additional contribution to the phases of the four semi-classical paths (LL, LR, RL, RR) that involves non-local interactions between particles that have no geometric interpretation. 
While this is highly contrived, we mention this possibility only to highlight that GIE does not provide any direct evidence of geometries in superposition beyond the Newtonian limit, and possible deviations beyond this limit are at least conceivable, such as in Refs.\cite{HallReginatto2018, DoenerGrossardt2022}. Indeed, some recent works highlight the possible difference between superposed geometries that have drastically  different curvatures and those that differ only by a displaced source mass, as in the GIE case. \cite{foo2021schrodinger, foo2022schr}

Overall, the various conclusions and possibilities for new physics are summarized in Figure \ref{fig:setup2}.

\subsection{Comparison to other works on interpretations of GIE}
Various recent works have considered the implications and interpretations of the GIE experiment. Here we briefly compare them to our results. 

While not directly addressing the GIE proposal, an interesting recent related line of research has emerged that focuses on Gedankenexperiments and what they imply for the quantization of gravity.\cite{Baym_2009, mari2016experiments, belenchia2018quantum, belenchia2019information,danielson2022gravitationally} Belenchia et al.\cite{belenchia2018quantum} show how a seeming paradox arises for the interferometric which-way information between two space-like separated masses, but that it can be avoided when quantizing gravitational radiation. Such arguments relate to GIE, as they operate in the same Newtonian limit and support the view of the need to quantize gravity. Nevertheless, the paradox in such Gedankenexperiments can also be resolved without quantized radiation, and thus the question of quantization from these arguments remains inconclusive.\cite{rydving2021gedanken,grossardt2021comment}

Shortly after the GIE proposals, Anastopoulos and Hu pointed out in Ref. \cite{AnastopoulosHu2018} that due to the static Newtonian limit in GIE, the physical mediators do not play a role and that the mediator-interpretation is misguided.
The authors argued that no quantization of the mediators can therefore be inferred from the Newtonian limit, \changes{and discussed it further in Refs. \cite{Anastopoulos_HU_2020CAT,Anastopoulos_Lagouvardos_savvidou_2021} }
This aligns with what we discuss in the present manuscript as well, mainly in section \ref{sec-EM} where we show that (unphysical) mediators only appear when a specific gauge choice is made. We make the additional observation that both local and non-local formulations are not gauge artefacts, but  viable interpretations of known relativistic physics.

In a recent work, Bose et al.\cite{bose2022mechanism} show how gravitational entanglement is explicitly generated in a local picture, in analogy with a previous work by Franson, \cite{Franson2011} as we also use in section \ref{sec:EM-Lorentz}. The toy model involves virtual mediators, albeit not the physical ones. However, this demonstration alone is insufficient to infer their existence, it merely shows that it is possible to describe entanglement in this form. One can equivalently describe the creation non-locally, as we show in Sec.\,\ref{sec:CoulombQEDFranson} and Appendix \ref{sec:AppendixQEDCoulomb}. This invalidates somewhat the claims in Ref.\cite{bose2022mechanism}, where it is argued that the presence of virtual gravitons is essential for the masses to get entangled. The same can be achieved non-locally without gravitons in this proposal, as we showed in Sec.\,\ref{sec-GR} and Appendix \ref{sec:Appendix_weak_field_poisson}.

Providing a complementary perspective on GIE and inferred quantization, Carney has proven the existence of quantized physical gravitons based on three assumptions:\cite{Carney2021} that quantized source particles interact via a $1/r$ potential in the non-relativistic limit, that the $S$ matrix theory is unitary and that it is Lorentz invariant.  This seems in contradiction with the discussion in our work here, which highlights  that a quantized absorber theory of weak-field gravity provides an alternative theory without gravitons. 
The resolution of this seeming contradiction is a loophole in Carney's proof. If one is willing to make the cosmological assumption that the far future contains a complete gravitational absorber, then the $1/r$ potential, the unitarity and the Lorentz invariance do not require the existence of physical graviton states, or a Hilbert space associated with such states.  This is discussed in more detail in Appendix \ref{sec:CarneyAbsorberRelation}.

It was pointed out by Rovelli and Christodoulou that the fundamental mediators of gravity are not the central objects of interest in the GIE setup. The authors instead highlight the quantum nature of the sources and  conclude that an expected outcome in the GIE experiment would imply that gravitational fields can exist in superposition.\cite{RovelliChristod2019} Superposition of geometries, as discussed in Sec.\,\ref{subsec:superposed_geometries} are then inferred from an extension of the Newtonian superpositions to general relativistic superpositions.

In a recent paper, Christodoulou et al.\cite{ChristodoulouEtal2022} analyze GIE starting from a relativistic description based on path integrals, in order to highlight the inherent locality of entanglement generation in the relativistic setting. But as we have discussed in Sections \ref{sec-EM} and\ref{sec-path}, relativistic Lorentz invariance does not necessarily mean that physics has to be local,
but that causality has to be preserved.  The local interpretation is only true for the formulation in the Lorentz gauge, which is what is exclusively used in Ref.\cite{ChristodoulouEtal2022}. The same can be described in terms of non-local processes in the path integral formulation, both in a Poisson gauge and in terms of absorber theory, as we showed in section \ref{sec-path}. While a modified setup that allows for retardation effects as proposed in Ref.\cite{ChristodoulouEtal2022} would involve local terms also in the Poisson gauge, the absorber non-local loophole still remains.

Finally, we mention that some alternative models have been proposed that do not involve the expected quantized Newtonian fields but that can still explain a positive outcome of the GIE experiment.\cite{HallReginatto2018,hall2021comment,DoenerGrossardt2022} These unorthodox semi-classical models are interesting test theories that warrant further study, at the very least as possible test theories against which the expected predictions can be probed and to design schemes that may distinguish them.

\section{Conclusions}

In this paper we have critically assessed possible conclusions that one can draw from observing gravitationally induced entanglement. We have shown that entanglement can be generated by quantized mediators, but that such an interpretation is not unique. The possibility of non-local generation of entanglement is also viable even within known relativistic physics (absorber theory being one example), and thus no conclusion can be drawn about mediators unless the locality condition is additionally imposed. While locality is a choice and not fundamental in both QED and general relativity, causality is always preserved even in non-local formulations. 
The experiment nevertheless probes quantum aspects of gravity \changes{independently of the LOCC argument}, as it involves superpositions of gravitational source masses, demonstrating the creation of superpositions of the Newtonian gravitational field. 

Any interpretation beyond this relies on assumptions about the nature of quantum gravity which may or may not be true, and thus remain ambiguous. \changes{If locality is elevated to a fundamental assumption, then LOCC indeed predicts the existence of quantized mediators if GIE is witnessed.} Intense debates about the proposal thus reach different conclusions, depending on what prior assumptions one is willing to accept as ``most natural''. Since the proposed tests are of indirect nature, the choice for the underlying ontology is crucial for the conclusions one can draw. The electromagnetic case differs somewhat from the gravitational one, as the former can always be cast in a non-local absorber formulation, while gravity is only known in the Newtonian and post-Newtonian limits to have a non-local absorber picture. In both cases, also an interpretation with local mediators of entanglement is permitted, although the involved mediators are not the observable physical bosons. But since non-local formulations are permitted as well, observing the expected entanglement would be compatible with both the non-existence and existence of local quantized mediators.
We also showed that under reasonable prior assumptions, one can already claim tests of the quantum nature of the Newtonian gravitational field in cosmological observations. Such a claim, however, relies additionally on the validity of the usual inflationary paradigm.

In sum, there are three types of conclusions one can draw from GIE: i) The experiment can probe some alternative theories that would yield an unexpected outcome, such as Penrose collapse, classical local channel models or mean field theory. ii) An expected result will show that spatial superpositions of matter coherently source Newtonian gravity, yielding superpositions of Newtonian fields. The caveat is that if one assumes the usual inflationary paradigm, matter in quantum states sourcing Newtonian gravity is already shown in current CMB observations. iii) If one assumes an underlying relativistic theory for gravity, different interpretations exist on how entanglement is generated. One of them includes local mediators, which however are not the usual observable bosons. In such an interpretation these mediators have to be quantized. But entanglement generation can equally be described in terms of non-local processes in the relativistic setting, thus induced entanglement is insufficient to conclude the quantization of mediators. 
Overall, our results clarify that while quantum superpositions of gravitational sources are probed in such proposals, claims about mediators remain ambiguous, both in the electromagnetic and in the gravitational setups. 

\section*{Acknowledgements}
We thank Erik Aurell, Fawad Hassan, Myungshik Kim, Jonas Larson, Florian Niedermann, Germaine Tobar, Frank Wilczek and Ting Yu for useful discussions and feedback. 
This work was supported by the Swedish Research
Council under grant no. 2019-05615, the European Research Council under grant no. 742104 and The Branco
Weiss Fellowship -- Society in Science.
\section*{DATA AVAILABILITY}
Data sharing is not applicable to this article as no new data were created or analyzed in this study.
\section*{AUTHOR DECLARATIONS}
\subsection*{Conflict of Interest}
The authors have no conflicts to disclose.
\subsection*{Authors contributions}
$^*$ VF and MK contributed equally to this work.
\section{Appendix}
\subsection{QED in Coulomb gauge}
\label{sec:AppendixQEDCoulomb}
For the sake of completeness, in this section we will explicitly show how entanglement between two charged quantum particles, is generated non-locally yet causally, in the Coulomb gauge. The interaction Hamiltonian between charged particles and EM field is 
\begin{equation}\label{QEDINTER}
    H_{int}=\int d^3r j_{\mu} A^{\mu}=\int d^3r\bigl( -\rho \Phi_{em}+\vec{j}\cdot\vec{A}\bigr)\,.
\end{equation}
Writing the electric field in terms of the potentials $\vec{E}=-\nabla \Phi_{em}-\partial_t\vec{A}$ and substituting into the Gauss law 
\begin{equation}
    \nabla\cdot\vec{E}=\frac{\rho}{\epsilon_0}
\end{equation}
we end up with a Poisson equation for  $\Phi_{em}$  
\begin{equation}\label{Constraint}
    \nabla^2 \Phi_{em}=-\nabla\cdot\frac{\partial \vec{A}}{\partial t}-\frac{\rho}{\epsilon_0}\,.
\end{equation}
The solution of \eqref{Constraint} is
\begin{equation}\label{scalarpot}
   \Phi_{em}(\vec{r})=\int d^3r'\frac{\nabla\cdot\partial_t\vec{A}+\frac{\rho}{\epsilon_0}}{4\pi|\vec{r}-\vec{r}'|}\,.
\end{equation}
From \eqref{scalarpot}, it's clear that the $0$ component of the four vector $A_{\mu}$, namely $\Phi_{em}$ , is not an independent degree of freedom. Instead, it's completely determined by $\vec{A}$ and the presence of matter $\rho$. It therefore means that the four vector $A_{\mu}$ contains three degrees of freedom. Moreover, once the $U(1)$ gauge symmetry of QED is fixed, we will end up with two physical degrees of freedom which correspond to the two polarizations of the photon. The Coulomb gauge condition reads
\begin{equation}\label{coulombgauge}
\nabla\cdot\vec{A}=0    \,.
\end{equation}
This condition is leaving behind 2 degrees of freedom and implies that the vector field $\vec{A}$ is transverse, namely $\vec{A}\equiv\vec{A}_{\perp}$. Relation \eqref{scalarpot} now takes the form 
\begin{equation}\label{coulombQED}
   \Phi_{em}(\vec{r})=\frac{1}{4\pi\epsilon_0}\int d^3r'\frac{\rho(\vec{r}')}{|\vec{r}-\vec{r}'|}
\end{equation}
which corresponds to the Coulomb potential.
The resulting interaction Hamiltonian in Coulomb gauge is therefore
\begin{equation}\label{QEDHamiltonian}
    H_{int}=-\frac{1}{8\pi \epsilon_0}\int d^3rd^3r'\frac{\rho(\vec{r})\rho(\vec{r}')}{|\vec{r}-\vec{r}'|}+\int d^3r\vec{j}\cdot\vec{A}_{\perp}\,.
\end{equation}
Upon quantization, this Hamiltonian is responsible for entangling two charged quantum particles via the non-local, instantaneous Coulomb interaction and/or via virtual spin-1 photons. In the non-relativistic regime in which the latter is irrelevant\cite{Franson2011}, namely for slowly moving particles, as it is the case for electromagnetic analogue of the GIE proposal,\cite{BoseAnupam, MarlettoVedral2017} we have to conclude that entanglement generation is due to the Coulomb term, which corresponds just to quantized source masses. This conclusion had also been pointed out in Ref. \cite{AnastopoulosHu2018}.

It's crucial to notice that despite the various non-local (instantaneous) pieces that appear in the formulation, QED in Coulomb gauge describes causal physical processes.\cite{Jackson2002,Wong2010, Jefimenko1989, Rohrlich2002, PhysRevA.38.4897}

\subsection{Weak-field gravity in Poisson gauge}
\label{sec:Appendix_weak_field_poisson}
The aim of this section is to show how entanglement between the masses can be generated non-locally without any reference to gravitons. We expand on the exposition in Ref. \cite{Bertschinger1999}. In the linear regime of general relativity, the metric tensor is written as  $g_{\mu\nu}=\eta_{\mu\nu}+h_{\mu\nu}$, where $|h_{\mu\nu}|\ll 1$ and $\eta_{\mu\nu}$ corresponds to the flat Minkowski metric tensor. Decomposing the linear field $h_{\mu\nu}$ according to its transformation properties under spatial rotations in terms of scalar, vector and tensor components $(\phi, \psi, w_i, s_{ij})$, we have 
\begin{align}
    h_{00}&=-2\phi  \\    h_{0i}&=w_i \\
    h_{ij}&=-2\psi\delta_{ij}+2s_{ij}
\end{align}
where $s_{ij}$ is the so-called traceless strain, $s_j^j=0$. So far, no gauge condition has been imposed. 
The interaction Hamiltonian $H_{int}$ between the weak gravitational field and matter is given 
 \begin{equation}
  H_{int}=-\frac{1}{2}\int d^3r h_{\mu\nu}(\vec{r})T^{\mu\nu}(\vec{r})
 \end{equation}
 or equivalently 
 \begin{equation}\label{ham}
  H_{int}=\int d^3r\bigl[\phi T^{00}+(\psi\delta_{ij}-s_{ij})T^{ij}-w_{(i}T^{i)0}\bigr]\,.
 \end{equation}
The Poisson gauge condition 
 \begin{align} \label{sTTdef}
      \partial_js^j_i=0 \quad \textrm{and} \quad
     \partial_iw^i=0
 \end{align}
implies that the traceless strain $s_{ij}$ as well as the vector $w_i$ are spatially transverse. \changes{An $s_{ij}$ that satisfies eq. \eqref{sTTdef} is denoted by $s^{TT}_{ij}$. Similarly we denote a $w_i$ satisfying eq. \eqref{sTTdef} by $w_i^{\perp}$.} In this gauge, the interaction Hamiltonian is written as 
\begin{equation}\label{interham}
H_{int}=\int d^3r\bigl[\phi T^{00}+(\psi\delta_{ij}-s_{ij}^{TT})T^{ij}-w_{(i}^{\perp}T^{i)0}\bigr]\,.
\end{equation}
The dynamics of remaining six degrees of freedom $(\phi,\psi,s_{ij}^{TT},w_i^{\perp})$ are dictated by 
Einstein's equation, which in the Poisson gauge takes the following form \cite{Bertschinger1999}
\begin{equation}\label{poi}
\nabla^2\psi = 4\pi G T_{00}
\end{equation}
\begin{equation}\label{Poisson}
    \nabla\cdot (-\nabla\phi-\partial_{t}\vec{w}_{\perp})-3\partial_t^2\psi=-4\pi G(T_{00}+T^i_i)
\end{equation}
\begin{equation}\label{amp}
    \nabla \times \vec{H}=-16\pi G \vec{f}_{\perp}
\end{equation}
\begin{align}\label{constraints}
    \bigl(\partial_i\partial_j-\frac{1}{3}\delta_{ij}\nabla^2\bigr)(\psi-\phi)&=8\pi G\Pi_{ij}^{||} 
\end{align}
\begin{equation}\label{GW}
    \Box s_{ij}^{TT}=8\pi G \Pi_{ij}^{TT}
\end{equation}
where $\vec{H}=\nabla\times \vec{w}_{\perp}$ and  $\vec{f}\equiv T^{0i}\vec{e}_i$. 

$\Pi^{TT}_{ij}$ corresponds to the transverse and traceless part of the stress energy tensor. Making use of the so-called Helmholtz decomposition, the three-vector $\vec{f}$ can be decomposed into longitudinal and transverse parts as follows
\begin{equation}
    \vec{f}=\vec{f}_{||}+\vec{f}_{\perp}
\end{equation}
where  
\begin{equation}
    \vec{f}_{\perp}=\frac{1}{4\pi}\nabla\times\nabla \cdot \int d^3r' \frac{\vec{f}(\vec{r}')}{|\vec{r}-\vec{r}'|}
\end{equation}
and
\begin{equation}
    \vec{f}_{||}=-\frac{1}{4\pi}\nabla \int d^3r'\frac{\nabla'\cdot\vec{f}(\vec{r}')}{|\vec{r}-\vec{r}'|} \,.
\end{equation}
The function $\Pi_{ij}$ is the traceless spatial part of the stress energy tensor
\begin{equation}
    \Pi_{ij}\equiv T_{ij}-\frac{1}{3}\delta_{ij}T^{k}_{k} \,.
\end{equation}
In a similar way, it can be decomposed as
\begin{equation}\label{helmholtztensor}
   \Pi_{ij} =\Pi_{ij}^{||}+\Pi_{ij}^{\perp}+\Pi_{ij}^{TT}
\end{equation}
where its longitudinal part is defined as
\begin{align}
    \nabla^2 \Pi_{ij}^{||}(\vec{r})&=\bigr(\partial_i\partial_j-\frac{1}{3}\delta_{ij}\nabla^2\bigl)\Pi_{||}(\vec{r}) \label{pilongit}\\
\Pi_{||}(\vec{r})&=-\frac{1}{4 \pi}\frac{3}{2}\int d^3 r' \frac{{\partial^i}'{\partial^j}' \Pi_{ij}(\vec{r}') }{|\vec{r} - \vec{r}'|}    \label{invertedpiparal}
\end{align}
and its rotational part 
\begin{equation}\label{pidiver}
    \Pi_{ij}^{\perp}(\vec{r})=\partial_i\Pi^{\perp}_j+\partial_j\Pi^{\perp}_i
\end{equation}
in terms of a divergence-less spatial vector $\Pi^{\perp}_j$. $\Pi^{\perp}_j$ in turn is defined in terms of $\partial^j \Pi_{ij}$ analogously to $\vec{f}_{\perp}$. The last term in \eqref{helmholtztensor}, $\Pi^{TT}_{ij}$, corresponds to the gauge invariant transverse part.
\changes{Applying the same steps to the metric perturbations provides a 
procedure for the extraction of $s_{ij}^{TT}$ from $h_{ij}$.}

We note in passing that, in analogy with electrodynamics, where the Poisson  (constraint) equation $\nabla\cdot\vec{E}=\rho/\epsilon_0$ enforces charge conservation $\partial_{\mu}J^{\mu}=0$, in the weak-field of GR,
gauge invariance, namely invariance under infinitesimal change of the coordinates $x^{\mu}\rightarrow x^{\mu}-\xi^{\mu}$, implies four conserved quantities $\partial_{\mu}T^{\mu\nu}=0$.
This equation shows the limitations of weak-field gravity since it contradicts the equations of motion of matter. This inconsistency can only be avoided after resumming all perturbation orders to get back full general relativity.

The set of equations \eqref{poi}-\eqref{GW} describe the causal dynamics of the metric components in Poisson gauge. Neglecting some pieces, or focusing only on the constraint equations, would erroneously lead to the conclusion that GR in the weak-field limit describes non-causal physics.

\begin{figure*}[t]
\centering
\begin{tikzpicture}[scale=.65,every node/.style={scale=0.65}]

\node[scale=1.5] at (-7,0) {\footnotesize Generic boundary condition:};

\draw (-2,-1) -- (-1,0);
\draw (-2,1) -- (-1,0);
\draw [dashed] (-1,0) -- (1,0); 
\draw (1,0) -- (2,1);
\draw (1,0) -- (2,-1);
\draw [photon] (-1.5,-2) -- (-1.5,-.5);
\draw [photon] (1.5,2) -- (1.5,.5);

\node at (-2,1.3) {${p}'_1$};
\node at (2,1.3) {${p}'_2$};
\node at (-2,-1.3) {${p}_1$};
\node at (2,-1.3) {${p}_2$};
\node at (-1.2,-2) {${k}$};
\node at (1.2,2) {${k}'$};
\node at (0,.3) {$\tilde{k}$};

\node[scale=1.5] at (3.8,0) {$\Rightarrow$};

\begin{scope}[xshift=230]

\draw (-2,-1.5) -- (-1,-.5);
\draw [photon] (0,-1.5) -- (-1,-.5);
\draw (-1,-.5) -- (-1,.5);
\draw (-1,.5) -- (-2,1.5);
\draw [dashed] (-1,.5) -- (0,1.5);

\draw (1,-1.5) -- (1,1.5);

\node at (-2,1.8) {${p}'_1$};
\node at (1,1.8) {${p}_2$};
\node at (-2,-1.8) {${p}_1$};
\node at (1,-1.8) {${p}_2$};
\node at (0,-1.8) {${k}$};
\node at (0,1.8) {$\tilde{{k}}$};

\end{scope}

\end{tikzpicture}


\centering
\begin{tikzpicture}[scale=.65,every node/.style={scale=0.65}]

\node[scale=1.5] at (-7,0) {\footnotesize Absorbing boundary condition:};
\draw (-2,-1) -- (-1,0);
\draw (-2,1) -- (-1,0);
\draw [dashed] (-1,0) -- (1,0); 
\draw (1,0) -- (2,1);
\draw (1,0) -- (2,-1);
\draw [photon] (-1.5,-2) -- (-1.5,-.5);
\draw [photon] (1.5,2) -- (1.5,.5);
\draw (2.75,-1.5) -- (2.75,1.5);

\node at (-2,1.3) {${p}'_1$};
\node at (2,1.3) {${p}'_2$};
\node at (-2,-1.3) {${p}_1$};
\node at (2,-1.3) {${p}_2$};
\node at (-1.2,-2) {${k}$};
\node at (1.2,2) {${k}'$};
\node at (0,.3) {$\tilde{k}$};
\node at (2.75,-1.7) {${p}_3$};
\node at (2.75,1.7) {${p}_3$};

\node[scale=1.5] at (4,0) {$\Rightarrow$};

\begin{scope}[xshift=230]

\draw (-2,-1.5) -- (-1,-.5);
\draw [photon] (0,-1.5) -- (-1,-.5);
\draw (-1,-.5) -- (-1,.5);
\draw (-1,.5) -- (-2,1.5);
\draw [dashed] (-1,.5) -- (0.25,1.25);

\draw (1.01,-1.5) -- (0.25,1.25);
\draw (0.25,1.25) -- (0.25,2.2);
\draw (2,-1.5) -- (2,1.5);

\node at (-2,1.8) {${p}'_1$};
\node at (2,1.8) {${p}_2$};
\node at (-2,-1.8) {${p}_1$};
\node at (2,-1.8) {${p}_2$};
\node at (-0.,-1.8) {${k}$};
\node at (-0.5,1.5) {$\tilde{{k}}$};
\node at (0.25,2.35) {${p}'_3$};
\node at (1.01,-1.8) {${p}_3$};

\end{scope}

\end{tikzpicture}

\caption{\small \textit{Upper panel}: Unitarity at the tree-level of a six-point amplitude (left) implies the existence of diagrams with on-shell graviton lines (right), as described in Ref.\cite{Carney2021}.
\textit{Lower panel}: Absorbing boundary conditions can be implemented by having one additional absorber particle $p_3$ with trivial propagator as disconnected line ``participating'' in the scattering on the left. In this case the pole can be removed from the scattering amplitude and the gravitons do not need to exist.}
\label{figure-treeunitarity}
\end{figure*}
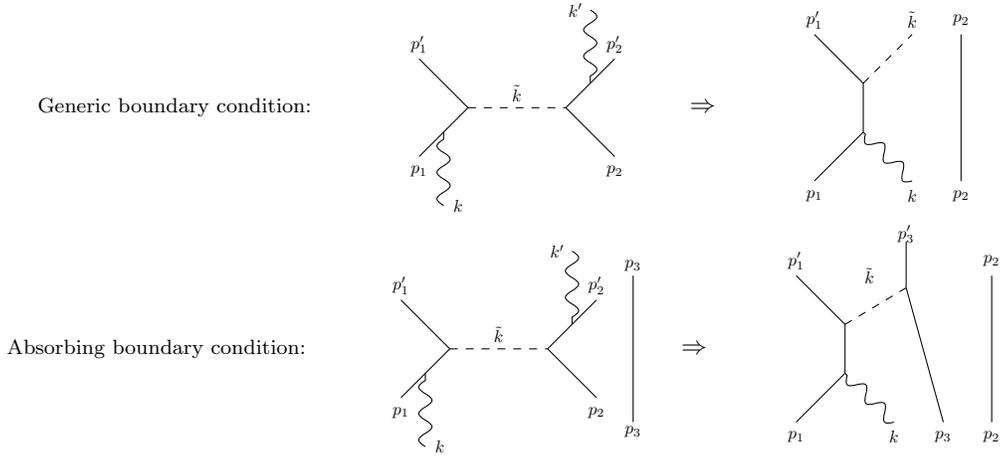

The solutions of \eqref{poi} and \eqref{amp} are easily obtained
\begin{equation}\label{psisol}
    \psi=-G\int d^3r'\frac{T_{00}(\vec{r}')}{|\vec{r}-\vec{r}'|}
\end{equation}
\begin{equation}\label{wsol}
    \vec{w}_{\perp}=-4G\int d^3r'\frac{\vec{f}_{\perp}(\vec{r}')}{|\vec{r}-\vec{r}'|}\,.
\end{equation}
Inserting  \eqref{pilongit} into $\eqref{constraints}$ one obtains the Poisson-like equation
\begin{equation}
\nabla ^2 (\psi-\phi)=8\pi G\Pi^{||}  \,
\end{equation}
which can be solved to give
\begin{equation}\label{poten}
    \phi(\vec{r})=\psi(\vec{r}) +2 G\int d^3r'\frac{\Pi^{||}(\vec{r}')}{|\vec{r}-\vec{r}'|}
\end{equation}
where the scalar function $\Pi_{||}(\vec{r})$ is defined in \eqref{invertedpiparal}.

Inserting the solution to constraint equations \eqref{psisol},\eqref{wsol} and \eqref{poten} into the  \eqref{interham} we end up with an expression for the interaction Hamiltonian which only depends on the gauge invariant spin-2 gravitons $s_{ij}^{TT}$ and the matter degrees of freedom
\begin{align}
H_{int}=& -\int d^3r\bigl(s_{ij}^{TT}(\vec{r})T^{ij}(\vec{r})\bigr)- \label{result} \\
         &
-\frac{G}{2}\int\frac{d^3rd^{3}r'}{|\vec{r}-\vec{r}'|}\biggl[ -4f_{\perp,i}(\vec{r}')T^{0i}(\vec{r})  + \notag\\
\quad &  +T_{00}(\vec{r}')\bigl(T^{00}(\vec{r})+T^{k}_{k}(\vec{r})-2\Pi^{||}(\vec{r}) \bigr) \biggr] \notag
\end{align}
where $f_{\perp,i}\propto T_{\perp}^{0i}$. The first term in \eqref{result} describes gravitational radiation. The second and third line correspond to non-local terms.  The third term contains the standard Newtonian instantaneous interaction between mass densities. Note that in the fully nonperturbative case the Hamiltonian can be decomposed in a similar fashion by reinserting the solution of algebraic and elliptic equations in a fully constrained gauge.\cite{BonazzolaGourgoulhonGrandclementEtal2004} Thus, slowly moving charges entangle
predominantly at short distances without the physical gravitational degrees of freedom even if spacetime curvature is strong. \changes{Quantization of the Hamiltonian \eqref{result} is straightforward, as all redundant gravitational degrees of freedom have already been removed and only the term $s_{ij}^{TT}$ remains, which corresponds to the spin-2 gravitons. Quantization then yields eq. \eqref{hamiltpoisson} in the main text. 
One can see that entanglement is generated either via the physical gravitons (1st term $\propto s_{ij}^{TT}$ in \eqref{result})) and/or via the non-local matter terms (the remaining terms).}

\changes{After quantization, and } since the experiment is quasi-static with all $v \ll c$, the only relevant component is the $\hat{T}_{00}(\vec{r})$ term. Therefore, \eqref{result} takes the form 
\begin{equation}
\hat{H}_{int} \approx   -\frac{G}{2}\int\frac{d^3rd^{3}r'}{|\vec{r}-\vec{r}'|} \hat{T}_{00}(\vec{r}')\hat{T}^{00}(\vec{r})
\end{equation}
For $\hat{T}_{00}(\vec{r})=m_1\delta(\vec{r}-\hat{\vec{r}}_1)+ m_2 \delta(\vec{r}-\hat{\vec{r}}_2)$, we get
\begin{equation}
\hat{H}_{int}=   -\frac{Gm_1 m_2}{|\hat{\vec{r}}_1-\hat{\vec{r}}_2|} 
\end{equation}
which corresponds to the Newtonian interaction, as discussed in Section \ref{sec-NR}.

\subsection{Graviton existence theorem and the absorber loophole}
\label{sec:CarneyAbsorberRelation}

It was recently proven in Ref.\cite{Carney2021} that unitarity and Lorentz invariance, together with the assumption that non-relativistically gravity is a $1/r$ potential, imply the existence of physical gravitons, or in other words the existence of quantum states of gravitational degrees of freedom. 
As we discussed in the main text, assuming locality and the LOCC argument does not directly allow to conclude the existence of physical graviton states, only the existence of auxiliary or ``unphysical'' gravitons that can only appear in virtual processes. These gravitons are the analog of the scalar and longitudinal photons of QED.

It therefore seems that Carney's existence proof of physical gravitons is not only stronger than the LOCC argument but also already shows the existence of the graviton without the need to perform the GIE experiment.
More strikingly, Carney's proof which is also formulated in the weak-field regime of gravity, seems to be in contradiction with the fact that an absorber theory of weak-field quantum gravity is unitary, Lorentz invariant and has a classical $1/r$ potential. 
Thus the properties that Carney assumed are present in an absorber theory, but that by construction has no gravitons!

In this appendix we point out the hidden additional assumption that goes into the proof and allows one
to infer the existence of the graviton, and that dropping this extra assumption is then consistent with the existence of a gravitational absorber theory that has no gravitons.
To summarise our result: unitarity and Lorentz invariance,  and the $1/r$ potential only imply the graviton if absorbing boundary conditions are excluded. This exclusion was the additional hidden assumption in Ref.\cite{Carney2021}

Carney considered a tree-level process in which two masses are controlled electromagnetically, for instance to produce the superpositions in the GIE experiment, and then interact through the $1/r$ potential.
Considering a six-point amplitude, see upper left diagram in Fig.\,\ref{figure-treeunitarity}, involving the photon (curvy lines), massive particles (full lines) and gravitational interaction (dashed line), the existence of diagrams with on-shell graviton lines is derived, see the upper right diagram. The pole at $\tilde{k}^2 \rightarrow 0$ has an imaginary residue. This residue is equal to the product of a pair of amplitudes, one shown on the upper right of Fig.\,\ref{figure-treeunitarity}, where the physical graviton  (dashed line) is emitted into the final state. 
The disconnected line represents a trivial propagator. If this diagram, and thus physical graviton states did not exist, unitarity is violated.
This was in a nutshell Carney's argument.

Now we come to the absorber loophole. In an absorber ontology the diagram in the upper left of Fig.\,\ref{figure-treeunitarity} is incomplete and misses the existence of absorber particles. While most absorber particles are not involved in any given scattering event, some always will be. This is indicated by the lower left diagram in Fig.\,\ref{figure-treeunitarity}.
This ``spectator'' absorber allows then to restore unitarity through the diagram shown in the lower right. It absorbs the would-be graviton. In the absorber ontology, there are no on-shell gravitons, so the graviton pole does not exist in the absorber theory, and thus restoration of unitarity only requires proper consideration of absorbing boundary condition, and no introduction of graviton quantum states.

Note that absorbing boundary conditions do not exclude the existence of a physical graviton state, but they do not require such a state either.
Thus gravitons cannot be inferred from Carney's argument if absorbing boundary conditions are imposed.

\def\aj{AJ}
\def\aap{A\&A}
\def\apj{ApJ}
\def\aapr{A\&A Rev.}
\def\apjl{ApJ}
\def\mnras{MNRAS}
\def\araa{ARA\&A}
\def\aj{AJ}
\def\pra{PRA}
\def\qjras{QJRAS}
\def\physrep{Phys. Rep.}
\def\nat{Nature}
\def\aaps{A\&A Supp.}
\def\apss{Ap\&SS}      
\def\apjs{ApJS}
\def\prd{Phys. Rev. D}
\def\prl{Phys. Rev. Letters}
\def\jcap{JCAP}
\def\nar{New Astron. Rev}
\bibliography{references.bib}
\end{document}